\documentclass[pra,twocolumn,superscriptaddress,amsmath,amssymb,floatfi,noshowpacs]{revtex4}
\usepackage{graphicx}

\begin{document}
\title{\textbf{Quantum optomechanics of a two-dimensional atomic array}}
\author{Ephraim Shahmoon}
\affiliation{Department of Physics, Harvard University, Cambridge, Massachusetts 02138, USA}
\author{Mikhail D.~Lukin}
\affiliation{Department of Physics, Harvard University, Cambridge, Massachusetts 02138, USA}
\author{Susanne F.~Yelin}
\affiliation{Department of Physics, Harvard University, Cambridge, Massachusetts 02138, USA}
\affiliation{Department of Physics, University of Connecticut, Storrs, Connecticut 06269, USA}
\date{\today}

\begin{abstract}
We demonstrate that a two-dimensional (2D) atomic array can be used as a novel platform for quantum optomechanics. Such arrays feature both nearly-perfect reflectivity and ultra-light mass, leading to significantly-enhanced optomechanical phenomena. Considering the collective atom-array motion under continuous laser illumination, we study the nonlinear optical response of the array.
We find that the spectrum of light scattered by the array develops multiple sidebands, corresponding to collective mechanical resonances, and exhibits nearly perfect quantum-noise squeezing. Possible extensions and applications for quantum nonlinear optomechanics are discussed.
\end{abstract}

 \maketitle

\section{Introduction}
The study of radiation pressure plays an important role in science and emerging technologies, from the manipulation of ions in quantum information processing \cite{CZ,MS}, to cooling and monitoring the motion of solid mirrors. \cite{AKM}.
These examples demonstrate the two extreme limits of light-induced motion, which are typically studied; namely, that of single atoms, and that of bulk objects. Situated in between these two extremes, this work deals with the optomechanics of a nearly-perfect mirror made of a single dilute layer of optically-trapped atoms.

It is well known that light can dramatically influence the motion of individual atoms, as demonstrated by laser-cooling of atoms \cite{CCT}. However, due to the small absorption cross-section of individual atoms, efficient optomechanical coupling typically requires interfacing light with highly reflective objects, such as optical cavities \cite{SK1,SK,ESS,CAM,RES}.
Most optomechanical systems involve the motion of bulk solid objects, such as a movable mirror or membrane inside a cavity, that are coupled to light via radiation pressure \cite{AKM,MEY,DOR,HAR}. While light can be strongly scattered in this way, its effect on the motion of such macroscopic objects is very limited, due to the extremely small zero-point motion of the latter. Although ground-state cooling of the mechanical state \cite{MAR,WIL,SCH,CHAN,TEU} and the generation of squeezed light \cite{HAM,SK3,REG,SAF} were recently achieved, reaching the single-photon optomechanical regime \cite{RAB,GIR} remains an outstanding challenge.

\begin{figure}
\begin{center}
\includegraphics[width=\columnwidth]{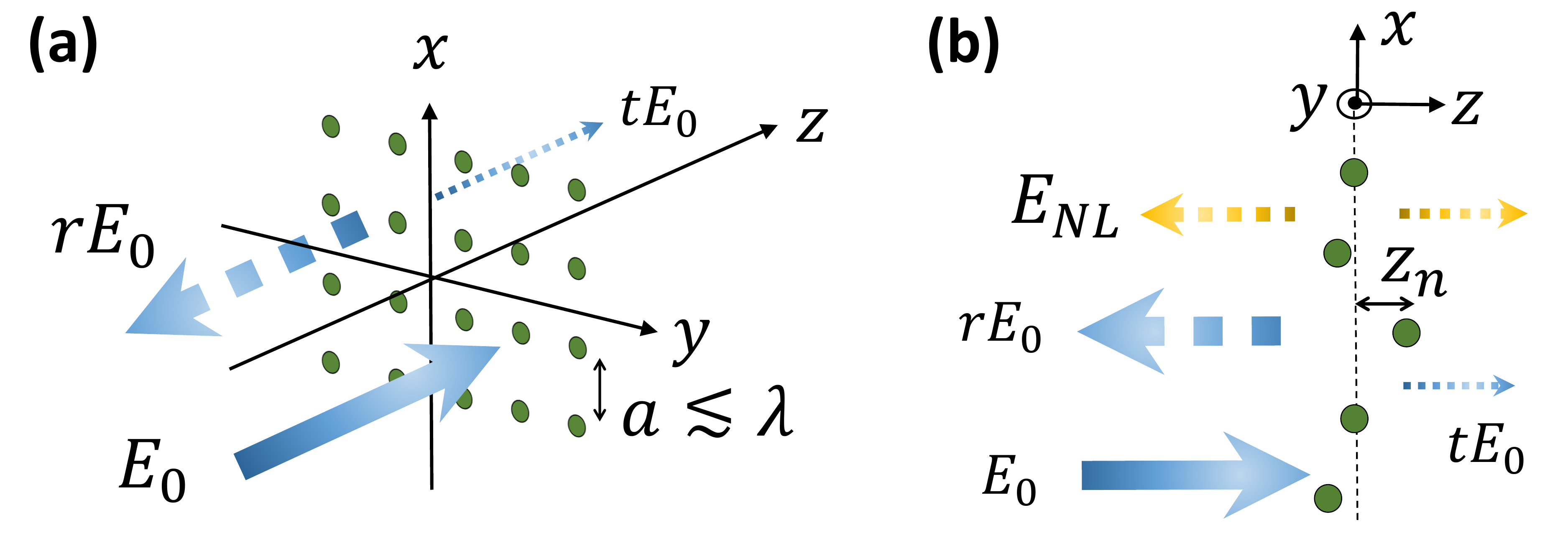}
\caption{\small{
Light scattering and optomechanics in an ordered 2D atomic array. (a) The atoms are spanning the $xy$ plane at equilibrium position $z=0$ for all atoms, with interatomic spacing $a$ on the order of the resonant wavelength of the atoms, $\lambda$. For non-saturated atoms (linear response), and ignoring their motion, full reflection is observed ($r=-1$, $t=0$) when the frequency of the incident light matches the cooperative resonance of the array \cite{coop}. (b) With longitudinal, light-induced atomic motion ($z_n$ for an atom $n$), a nonlinear component ($E_{\mathrm{NL}}$) is added to the reflected field, due to the optomechanical coupling.
 }} \label{fig1}
\end{center}
\end{figure}

In this work, we explore the optomechanics of a single 2D ordered array of optically-trapped atoms, as can be realized e.g. in optical lattices, in a cavity-free environment. It was recently shown, that such a 2D atom array can act as a nearly-perfect mirror, for light whose frequency matches the cooperative dipolar resonance supported by the array \cite{coop,ADM}. The mirror formed by such an array is easily pushed by the reflected light. Its zero-point motion is set by the depth of the atomic traps, which even for tight trapping (Lamb-Dicke regime), becomes $10^{-8}$m  to $10^{-7}$m, much larger than the $10^{-15}$m  to $10^{-13}$m zero-point motion of suspended bulk  mirrors or membranes \cite{AKM,HAR,BAC}.
Therefore, by combining nearly-perfect reflectivity with a high mechanical susceptibility, 2D atomic arrays could lead to very large optomechanical couplings.

We use a quantum-mechanical treatment to study the motion of atoms close to their equilibrium trap positions, under a continuous-wave laser illumination, which is weak enough to neglect internal-state saturation (Fig. 1). Cooperative effects due to dipole-dipole interactions play a central role in this system. First, they lead to a collective dipolar resonance of the \emph{internal} state of the atoms; and second, laser-induced dipolar forces between atoms lead to the formation of collective \emph{mechanical} modes. We show that the light-induced motion of this cavity-free many-atom system can be characterized by its mapping to a standard cavity optomechanics model in its bad-cavity, unresolved sideband regime. We then consider the back-action of this motion on the light, due to the optomechanical response of the array. In particular, we find that the collective mechanical modes imprint multiple sidebands on the spectrum of the light scattered by the array, and that this output light contains quantum correlations both in space and time, exhibiting large spatio-temporal squeezing.

These results provide a promising starting point and benchmark for further studies of optomechanics using ordered arrays of trapped atoms. They reveal that significant optomechanical couplings are achievable already at the level of a ``bare", cavity-free system of a single 2D array of dozens of atoms. More elaborate schemes may therefore enable reaching novel regimes of nonlinear and few-photon quantum optomechanics, as discussed below.

The article is organized as follows. Our theory of optomechanics of 2D atom arrays is presented in Secs. II and III. This includes the description of the system and its collective motion induced by light (Sec. II), and the characterization of the atom array system, via its mapping to the standard cavity optomechanical model (Sec. III). The theory is then applied to predict nonlinear optical phenomena, resulting from light-induced atomic motion: Sec. IV presents the analysis of the intensity spectrum of the output light, whereas Sec. V studies its quantum noise and correlation properties. Finally, we discuss some conclusions and future prospects in Sec. VI.

\section{Light-induced collective motion}
We consider a 2D array of trapped atoms $n=1,...,N$ at positions $\hat{\mathbf{r}}_n=(\mathbf{r}_n^{\bot},\hat{z}_n)$, illuminated by a right-propagating continuous-wave laser (Fig. 1). Motion is considered only along the longitudinal axis $z$, with $\hat{z}_n$ around the equilibrium position $z=0$, whereas the transverse positions $\mathbf{r}_n^{\bot}$ are assumed to be fixed (deep transverse trapping), forming a 2D lattice in the $xy$ space, e.g. a square lattice with lattice spacing $a$. Our theory below assumes an infinite array, but in practice it is valid for finite mesoscopic arrays ($\sqrt{N}\gg 1$, e.g. $N\sim10^2$) \cite{notes}.
The atoms are modelled as two-level systems with transition frequency $\omega_a$ and radiative width $\gamma$. Dipolar interactions between the array atoms, however, lead to a cooperative shift $\Delta$ and width $\Gamma$ of the atomic transition, reflecting the fact that the atomic dipoles respond collectively to light \cite{coop}. Nevertheless, for our purposes, these collective dipole modes effectively behave as individual atoms with a ``renormalized" (cooperative) resonance frequency $\omega_a+\Delta$ and width $\gamma+\Gamma$ \cite{notes}.
In the following, we discuss the light-induced collective motion of the array atoms. This discussion derives largely from Ref. \cite{notes}, briefly reviewed in Appendix A.

\begin{figure}
\begin{center}
\includegraphics[scale=0.27]{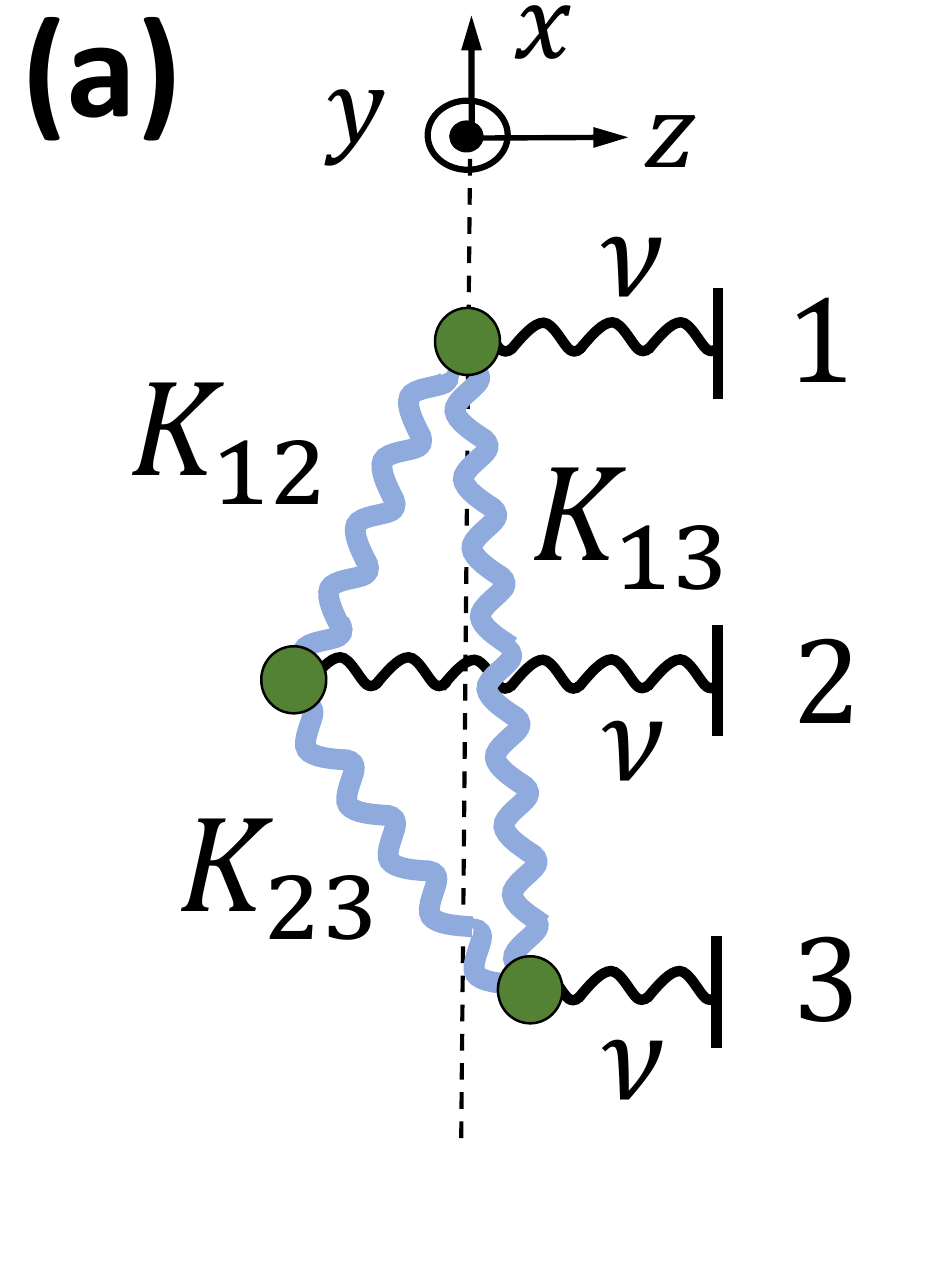}
\includegraphics[scale=0.035]{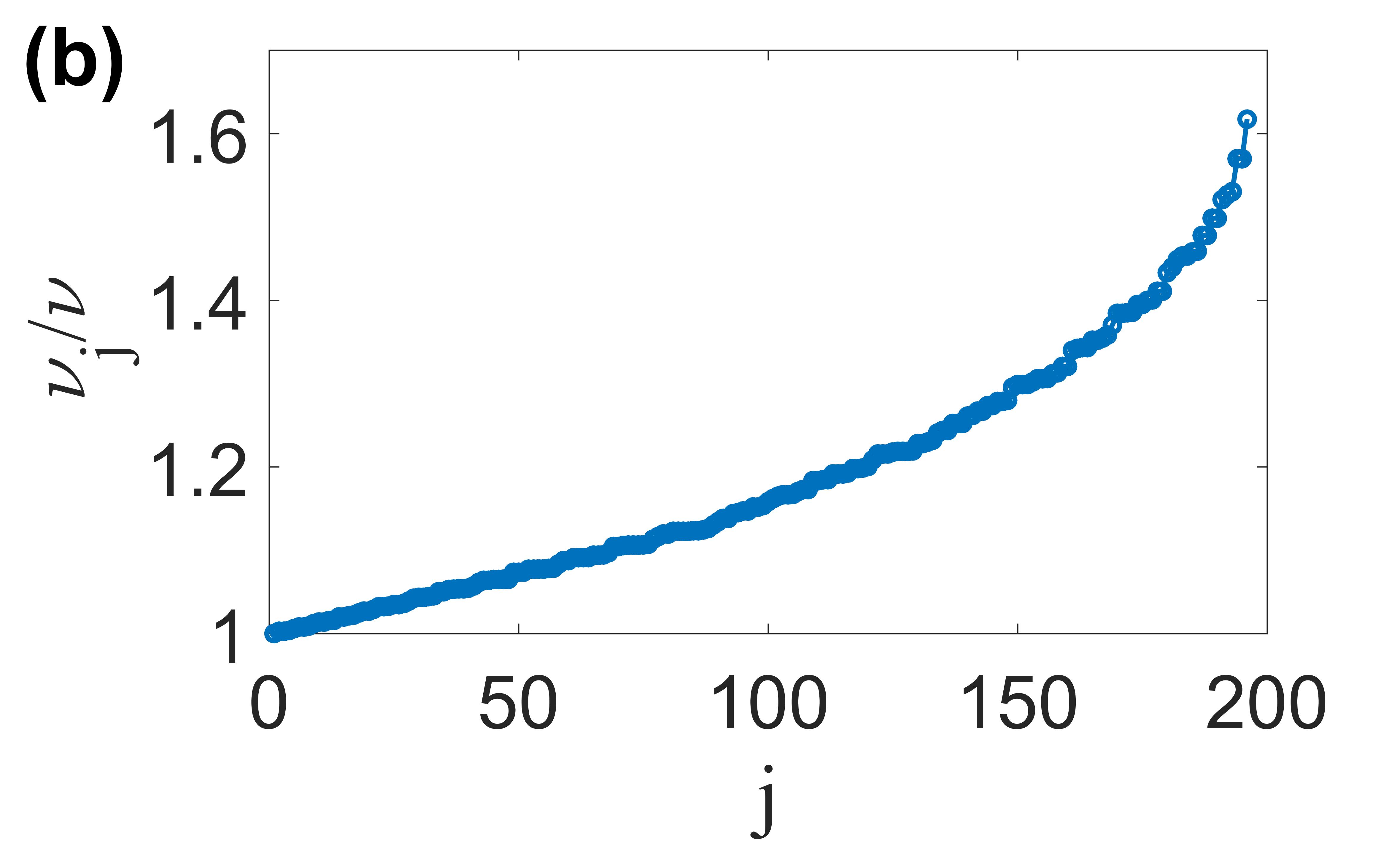}
\includegraphics[scale=0.035]{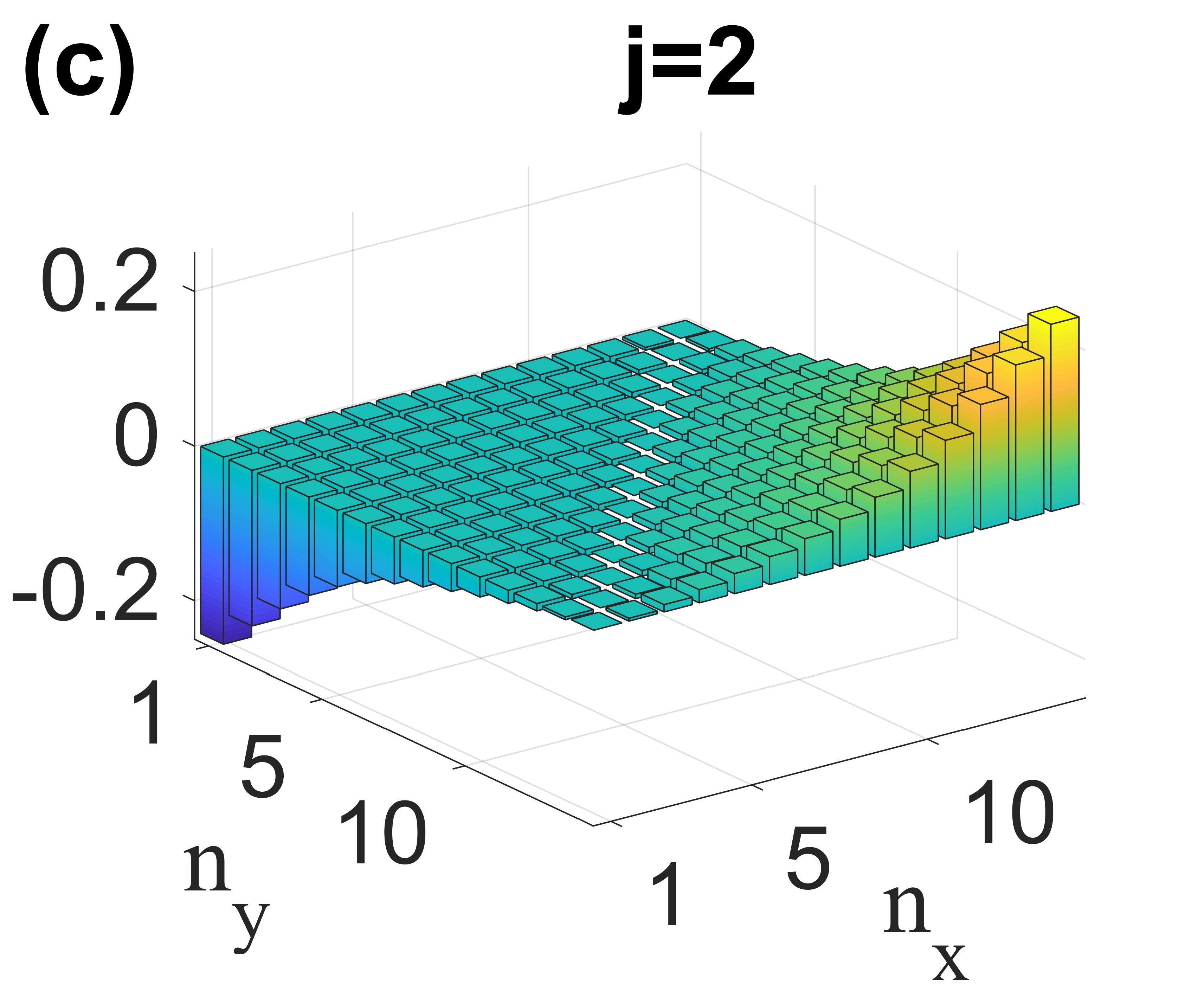}
\includegraphics[scale=0.035]{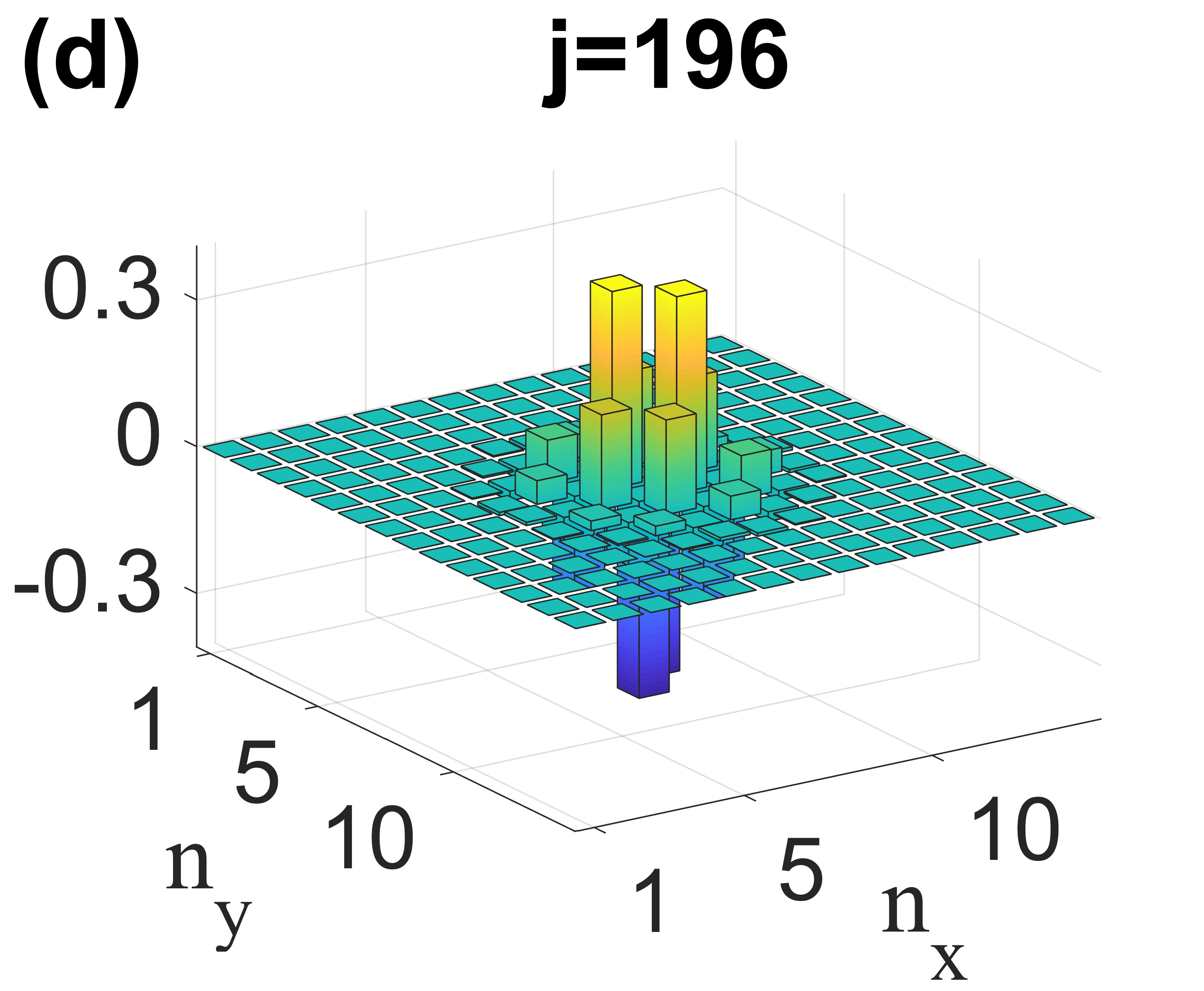}
\caption{\small{
Collective mechanical response of the array. (a) Coupled mechanical oscillators model corresponding to Eq. (\ref{EOM}): each atom is found in a potential with trap frequency $\nu$ (thin black ``springs") and coupled via laser-induced dipolar interactions $K_{nm}$ to the rest of the atoms (thick blue ``springs" connecting different atoms). (b) Eigenfrequencies $\nu_j$ of the resulting mechanical collective normal modes $j=1,...196$, for $N=14^2$ atoms, lattice constant $a/\lambda=0.2$ and detuning $\delta_L-\Delta=-(\gamma+\Gamma)/4$. (c,d): spatial profile of the mechanical collective modes $j=2$ (c) and $j=196$ (d) from the example in (b). The incident beam, with waist $w_0/\lambda=1.5$, is smaller than the array; therefore the profile of the highest frequency mode $j=196$ is highly oscillatory around the center of the array, where the beam intensity and the interactions it induces are strongest.
Other physical parameters: incident beam strength at the center $\Omega=\gamma$ ($\Omega\ll \gamma+\Gamma$), recoil energy $E_R/(\hbar\gamma)=1/810$ (corresponding to $^{87}$Rb), and Lamb-Dicke parameter $\eta=q x_0=0.12$  (corresponding to e.g. a potential depth $V=1000E_R$, trap length $l=450$nm and wavelength $\lambda=780$nm).
 }} \label{fig2}
\end{center}
\end{figure}

The derivation of the governing equation of atomic motion is based on the following considerations.
First, we take advantage of the separation of timescales between the fast internal and slow external atomic degrees of freedom, given by the cooperative decay rate $\gamma+\Gamma=\gamma\frac{3}{4\pi}(\lambda^2/a^2)$ \cite{coop} and the recoil energy $E_R=\hbar^2q^2/m$, respectively ($q=\omega_L/c=2\pi/\lambda$ being the laser wavenumber and $m$ the atom mass). This allows to adiabatically eliminate the internal degrees of freedom, obtaining a dynamical equation for the external, motional degrees of freedom $\hat{z}_n$. Second, we assume that the atoms remain inside the optical traps of length $<\lambda$, allowing to approximate $|\hat{z}_n|\ll \lambda$. Considering also atoms far from saturation (linearly responding, $\gamma+\Gamma\gg \Omega$, $\Omega$ being the Rabi frequency), we finally obtain (Appendix A):
\begin{eqnarray}
\dot{\hat{p}}_n&=&-m\nu^2\hat{z}_n+\bar{f}_n-\alpha_n\hat{p}_n+\hat{f}_n(t)+\sum_{m \neq n} K_{nm}(\hat{z}_n-\hat{z}_m),
\nonumber\\
\dot{\hat{z}}_n&=&\hat{p}_n/m,
\label{EOM}
\end{eqnarray}
with $\hat{p}_n$ the momentum of atom $n$.
This equation describes a collective Brownian motion, with the explicit expressions for the coefficients  $\bar{f}_n,\alpha_n,\hat{f}_n(t),K_{nm}$ given in Appendix A. The first term in Eq. (\ref{EOM}) is the restoring force due to the individual trap of an atom (longitudinal trap frequency $\nu$), whereas the next three terms account for light-induced forces including the average force $\bar{f}_n$, and the scattering-induced friction $\alpha_n$ and corresponding Langevin force $\hat{f}_n(t)$. The expressions for $\bar{f}_n$, $\alpha_n$ and $\hat{f}_n(t)$ resemble those from known single-atom theories of light-induced motion \cite{CCT}, except that here the atom-laser detuning $\delta_L$ and width $\gamma$ are modified by their cooperative counterparts $\delta_L-\Delta$ and  $\gamma+\Gamma$, respectively.

The term with coefficient $K_{nm}$ gives rise to a mechanical coupling between the atoms originating in the laser-induced dipole-dipole forces between pairs of atoms \cite{LIDDI}. It reflects that the motion of individual atoms is not independent, resulting in collective mechanical modes. Since $K_{nm}\propto \Omega_n^{\ast}\Omega_m$, with $\Omega_n$ the Rabi frequency on atom $n$, the collective mechanical modes crucially depend on the spatial profile of the incident light. To find the modes, we diagonalize Eq. (\ref{EOM}) in the absence of forces $\bar{f}_n,\hat{f}_n$ and friction $\alpha_n$, which amounts to the system of coupled oscillators from Fig. 2a. The collective mechanical normal modes of a square array with $N=14^2$ atoms, illuminated by a normal-incident Gaussian beam with waist smaller than the array size, are shown in Fig. 2b (eigenfrequencies) and Figs. 2c,d (spatial profiles).

For times $t$ longer than $1/\alpha_n$, the atomic motion in frequency domain becomes (Appendix A),
\begin{eqnarray}
\hat{z}_n(\omega)&=&\sum_j U_{jn} \hat{z}_j(\omega),
\nonumber\\
\hat{z}_j(\omega)&=&\bar{z}_j 2\pi\delta(\omega)+\frac{1}{m\nu_j^2}\chi_j(\omega)\hat{f}_j(\omega),
\nonumber\\
\chi_j(\omega)&=&-\frac{\nu_j^2}{\omega^2-\nu_j^2+i\alpha_j\omega}, \quad \bar{z}_j=\frac{\bar{f}_j}{m \nu_j^2}.
\label{zo}
\end{eqnarray}
Here $U_{jn}$ is the matrix element of the unitary transformation from the real-space lattice basis $n$ to the collective normal mode basis $j$ with eigenfrequencies $\nu_j$, $X_j=\sum_n U^{\ast}_{jn} X_n$ for $X=\bar{f},\hat{f},\hat{z}$ and $\alpha_j=\sum_j |U_{jn}|^2\alpha_n$. The solution $\hat{z}_j(\omega)$ for each normal mechanical mode $j$ consists of an average static shift $\bar{z}_j$ due to the static force $\bar{f}_n$ and a fluctuating part due to the linear mechanical response $\chi_j(\omega)$ to the corresponding Langevin force $\hat{f}_j(\omega)$.

Throughout this work, we assume that the atoms remain trapped, requiring that the potential depth $V$ of the traps is larger than the effective temperature $T_e$ associated with the Langevin force (Appendix A),
\begin{equation}
T_e=\frac{\hbar\gamma}{2}\frac{(\delta_L-\Delta)^2+\left(\frac{\gamma+\Gamma}{2}\right)^2}{(\Delta-\delta_L)(\gamma+\Gamma)}.
\label{Te}
\end{equation}
We note that for the atoms to remain trapped, $T_e$ has to be positive (and lower than the trapping potential), leading to the requirement of red cooperative detuning, $\delta_L<\Delta$.

\section{Mapping to cavity optomechanics}
Typical optomechanical systems can be modeled by a single optical cavity (boson mode $\hat{c}$) whose resonant frequency linearly depends on the position of a moving mirror (coordinate $\hat{z}\propto\hat{b}+\hat{b}^{\dag}$), as depicted in Fig. 3a, and with the Hamiltonian
\begin{equation}
H=\hbar\omega_c\hat{c}^{\dag}\hat{c}+\hbar\nu \hat{b}^{\dag}\hat{b}+\hbar g\hat{c}^{\dag}\hat{c}(\hat{b}+\hat{b}^{\dag})+(\hat{c}^{\dag}\Omega e^{-i\omega_L t}+\mathrm{h.c.}),
\label{Hc}
\end{equation}
$g$ being the bare optomechanical coupling and $\Omega$ the input field. It is therefore instructive to relate this simple standard model to the optomechanics of the atom array: Although the latter system does not include an optical cavity, it does include a resonator in the form of the internal degrees of freedom of the atoms.
To this end, we consider the linearized regime of the cavity optomechanics model, wherein the quantum fluctuations in the cavity and the motion are assumed to be much smaller than their corresponding classical steady-state values, and the Hamiltonian in a laser-rotated frame becomes \cite{AKM,MEY}
\begin{equation}
H\approx -\hbar\delta_c\hat{c}^{\dag}\hat{c}+\hbar\nu \hat{b}^{\dag}\hat{b}+\hbar (\bar{g}^{\ast}\hat{c}+\bar{g}\hat{c}^{\dag})(\hat{b}+\hat{b}^{\dag}),
\label{Hcl}
\end{equation}
where $\delta_c$ is a shifted laser-cavity detuning \cite{AKM}, $\bar{g}=g\bar{c}$ with $\bar{c}$ the classical steady-state value of the cavity field (c-number), and $\hat{c}$ and $\hat{b}$ the quantum fluctuations of the field and motion, respectively.

\begin{figure}
\begin{center}
\includegraphics[width=\columnwidth]{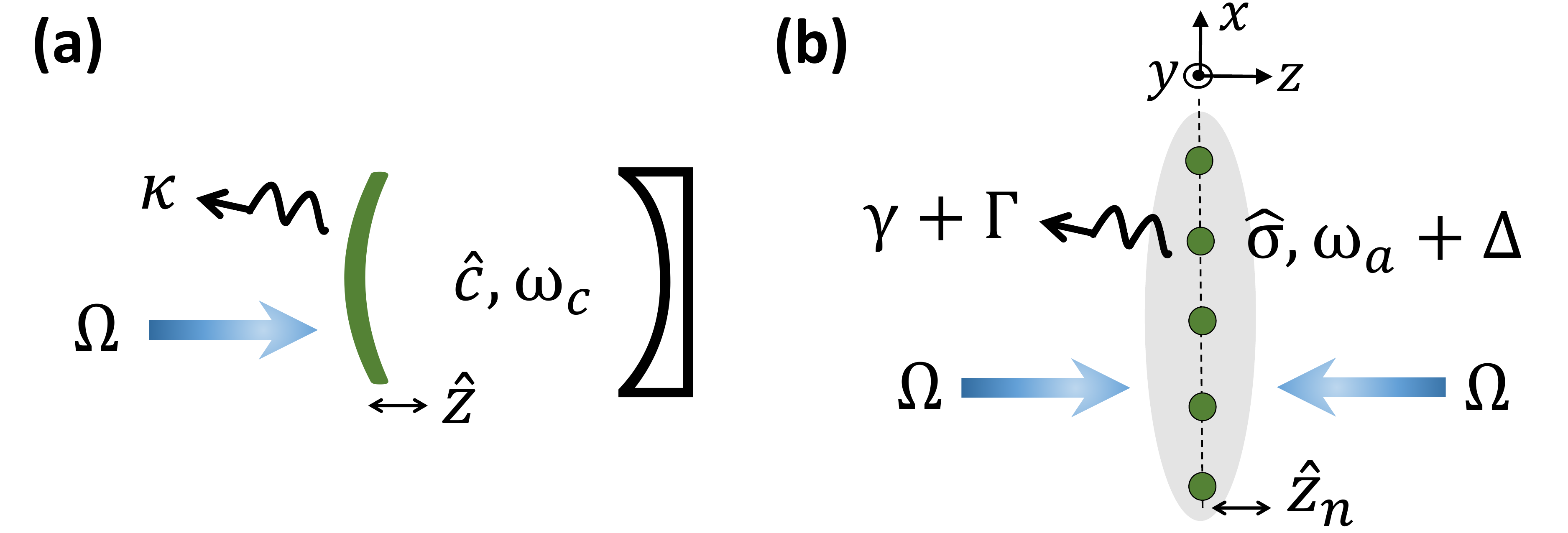}
\caption{\small{
Comparison with cavity optomechanics. (a) Standard cavity optomechanics model, Eq. (\ref{Hc}): a cavity mode $\hat{c}$ (resonant frequency $\omega_c$ and width $\kappa$) is formed by two mirrors, one of which movable (coordinate $\hat{z}$). (b) 2D atom array: dipole-dipole interactions between the atoms form a collective dipole of the array atoms ($\hat{\sigma}$) with a resonant frequency shifted by $\Delta$ and widened by $\Gamma$ from the single atom resonance $\omega_a$ and $\gamma$. Longitudinal motion of an atom $n$ inside the trap is described by the coordinate $\hat{z}_n$. The analogy between the systems is captured by the mapping from table I.
 }} \label{fig3}
\end{center}
\end{figure}

In contrast, consider now a light-matter interaction Hamiltonian, such as that used to derive Eq. (\ref{EOM}), $H_{\mathrm{int}}\sim \hat{\sigma}_n^{\dag} \hbar\Omega_n e^{iq \hat{z}_n}+\mathrm{h.c.}$, with $\hat{\sigma}_n$ the atomic-transition lowering operator and $\Omega_n$ the Rabi frequency at atom $n$. For $|\hat{z}_n|\ll \lambda$, its relevant optomechanical coupling becomes
\begin{equation}
H_{\mathrm{int}}\sim \hbar\hat{\sigma}_n^{\dag} \Omega_n q\hat{z}_n+\mathrm{h.c.}=\hbar(\eta\Omega_n^{\ast}\hat{\sigma}_n+\mathrm{h.c.})(\hat{b}_n+\hat{b}_n^{\dag}),
\label{Hint}
\end{equation}
where $\eta=q x_0$ is the Lamb-Dicke parameter, with $x_0=\sqrt{\hbar/(2m\nu)}$ the zero-point motion of an atom inside the trap. The form of $H_{\mathrm{int}}$ is identical to that of the interaction term in (\ref{Hcl}), with the internal, dipolar resonances of the linearly-responding atoms in the former, replacing the optical cavity resonator in the latter.
Focusing, for now, on a single motional degree of freedom of the array (e.g. a single atom), this suggests the following mapping between the 2D atom array and the cavity optomechanics models (Fig. 3b):
\begin{table}[ht]
\centering % used for centering table
\begin{tabular}{c|c|c} % centered columns (4 columns)
\hline\hline %inserts double horizontal lines
   & \textbf{cavity model} & \textbf{atom array} \\ [0.5ex] % inserts table
%heading
\hline % inserts single horizontal line
      resonator mode & $\hat{c}$ & $\hat{\sigma}$\\
      optomechanical coupling & $\bar{g}$ & $\eta \Omega$\\
      laser detuning & $\delta_c$ & $\delta_L-\Delta$ \\
       resonator damping rate & $\kappa$ & $\gamma+\Gamma$\\ [1ex] % [1ex] adds vertical space
\hline %inserts single line
\end{tabular}
\caption{Mapping to cavity optomechanics} % caption of Table
\label{table} % is used to refer this table in the text
\end{table}

Here, we recall the renormalized (cooperative) resonance of the array atoms with frequency $\omega_a+\Delta$ and width $\gamma+\Gamma$ ($\delta_L=\omega_L-\omega_a$ being the ``bare" laser-atom detuning).

More formally, the above mapping can be justified by deriving the equations of motion for the coordinate $\hat{z}$ and resonator $\hat{c}$ of the standard cavity model, and comparing them with the analogous equations for $\hat{z}_n$ and $\hat{\sigma}_n$ of the atom array. In Appendix B, we show that these two sets of equations are indeed equivalent, by considering the mapping from table I. Moreover, for the specific case of a bad cavity in the weak-coupling and unresolved sideband regimes, $\kappa\gg \bar{g},\nu$, the resonator mode $\hat{c}$ can be adiabatically eliminated, and the resulting equation of motion for $\hat{z}$ is essentially identical to Eq. (\ref{EOM}), for a single atom.

The multimode, many-atom collective mechanics, i.e. including the term $K_{nm}$ in Eq. (\ref{EOM}), can also be captured by the cavity model: It requires to include multiple mechanical modes in the Hamiltonian (\ref{Hc}) via an interaction term $\hbar\sum_n g_n\hat{c}^{\dag}\hat{c}(\hat{b}_n+\hat{b}_n^{\dag})$, resulting in an effective coupling parameter $K'_{nm}\propto \bar{g}_n \bar{g}_m$ (Appendix B). In contrast to the cavity model however, the multimode character of the 2D atom  array also extends to the output field, resulting in qualitatively new features as explored in the following.

\section{Mechanical sidebands in output light}
We now turn to study the optomechanical backaction on the light, in the form of nonlinear optical phenomena.
For non-saturated atoms, for which the polarizability is linear, optical nonlinearity originates only in the motion, via the following mechanism: The light pushes the atoms, whose positions are then determined by the
intensity of light. In turn, the phase of the light that is scattered off the atoms depends on their positions. This leads to an intensity-dependent phase, as in an optical Kerr medium. More formally, the reflected field from an atomic array is given by the scattered fields from all atoms, each of which is proportional
to a phase factor $e^{i 2q z_n}$. For an incident field $E$, radiation pressure leads to $z_n\propto |E|^2$ and hence to
intensity-dependent phase factors.

In this section we show that the multimode nature of the atom array optomechanics discussed above, manifests itself in the form of sidebands in the spectrum of the output light. The sidebands are located at the resonant frequencies $\nu_j$ of the collective mechanical modes $j$ at which the motion $\hat{z}_n$, and hence the phase factors $e^{i 2q z_n}$, are modulated; and the corresponding weights of these sidebands depend on the spatial profiles of these modes.

\subsection{Output light and nonlinearity}
The field scattered off an array of atoms has the form $\sum_n e^{-ik_z \hat{z}_n} \hat{\sigma}_n$. Using the adiabatic solution for the linearly-responding atomic dipoles, $\hat{\sigma}_n(t)$, we obtain the output field in the paraxial approximation (Appendix C)
\begin{eqnarray}
\widetilde{a}_{\mathbf{k}_{\bot}ks}&=&\beta_{\mathbf{k}_{\bot}}\delta_{s,+}\delta_{kq}+\hat{a}_{\mathbf{k}_{\bot}ks}
-g^{\ast}_0 \int_{-\infty}^{\infty}dt e^{i(ck-\omega_L)t}
\nonumber\\
&&\times\sum_n e^{-i\mathbf{k}_{\bot}\cdot\mathbf{r}_{n}^{\bot}}\sum_{s'=\pm}e^{-i(s-s')q\hat{z}_n(t)}\frac{\hat{\Omega}_{ns'}(t)}{\delta_L-\Delta+i\frac{\gamma+\Gamma}{2}}.
\nonumber\\
\label{aout}
\end{eqnarray}
Here, the ``output field", $\widetilde{a}_{\mathbf{k}_{\bot}ks}\equiv\hat{a}_{\mathbf{k}_{\bot}ks}(t=\tau) e^{i kc \tau}$, is the slow envelope of the lowering operator of the right/left-propagating ($s \rightarrow \pm$) photon mode with wavevector $\mathbf{k}=(\mathbf{k}_{\bot},k_z=sk)$ ($k\gg|\mathbf{k}_{\bot}|)$, evaluated at the final time $\tau\rightarrow \infty$, much after the atom-laser interaction ends. The ``input fields", $\hat{a}_{\mathbf{k}_{\bot}ks}\equiv \hat{a}_{\mathbf{k}_{\bot}ks}(t=-\tau) e^{-i kc \tau}$, are in turn evaluated at the initial time $-\tau\rightarrow -\infty$ before the atom-laser interaction begins, and are hence equal to vacuum fields satisfying, $\hat{a}_{\mathbf{k}_{\bot}ks}|0\rangle=0$. The coherent laser input is represented by the average amplitude (c-number) $\beta_{\mathbf{k}_{\bot}}=(1/N)\sum_n e^{-i\mathbf{k}_{\bot}\cdot\mathbf{r}_{n}^{\bot}}\beta_{n}$, which is related to the Rabi frequency via $\beta_{n}=-i\Omega_{n}/g_0$, with $g_0=\sqrt{\omega_L/(2\varepsilon_0\hbar A L)}d$ the atom-field coupling in the paraxial approximation ($d$ is the dipole matrix element, and $A$ and $L$ the quantization area and length). The Kronecker deltas $\delta_{s,+}$ and $\delta_{kq}$ represent a right-moving laser with frequency $\omega_L=qc$, and $\hat{\Omega}_{ns'}$ is the total Rabi frequency operator (including the input vacuum fluctuations) of the right/left-propagating ($s'\rightarrow \pm$) incident field.

In the absence of motion, $\hat{z}_n\rightarrow 0$, the output field is that due to the mirror-like linear response of the ordered atom array \cite{coop,ADM} (Appendix C). Frequency components other than that of the laser, appear in the output field due to the motion-induced phase factors $e^{-i(s-s')q\hat{z}_n(t)}$, and originate in a nonlinear optomechanical effect, $\hat{z}_n(t)\sim \hat{f}_n(t)\sim \hat{\Omega}_n^{\dag}(t)\hat{\Omega}_n(t)$.
They are most dominant in the \emph{reflected} field, since the phase factors exist only for $s\neq s'$ (oppositely-propagating input and output). We note that this analysis is valid only for the paraxial part of the output field, and can be therefore understood by considering the energy-momentum conservation of a photon colliding with an atom in 1D, where a forward-scattered photon cannot change its energy \cite{comment1}.

\subsection{Intensity spectrum of reflected light}
Consider now the detection of the left-propagating output field $s\rightarrow-$. Its dominant, average component is the linear reflection of the normal incident laser ($\mathbf{k}_{\bot}\approx 0$). In addition,
there exist nonlinearly-scattered field fluctuations at various transverse wavevectors $|\mathbf{k}_{\bot}|>0$, which can be detected at the corresponding far-field angles (Fig. 4b). These components originate in fluctuations in $\hat{z}_n$ which result in an effectively disordered array and therefore in scattering angles beyond that of a flat mirror.
The relevant spectrum of this detected total field is defined by
\begin{eqnarray}
I_{\mathbf{k}_{\bot}}(\omega)&=&\langle \widetilde{a}_{\mathbf{k}_{\bot}k-}^{\dag} \widetilde{a}_{\mathbf{k}_{\bot}k-}\rangle\frac{1}{|\beta_{\mathbf{k}_{\bot}=0}|^2}\frac{L}{c},
\nonumber\\
\label{Idef}
\end{eqnarray}
where $\omega=ck-\omega_L$ is the frequency of the field envelope around $\omega_L$, and the averaging is performed with respect to the field vacuum $|0\rangle$. The normalization is with respect to the dominant $\mathbf{k}_{\bot}=0$ component of the normal-incident field $\beta_{\mathbf{k}_{\bot}}$, and with $L/c\rightarrow 2\tau$ (experiment time). We note that this definition coincides with the standard definition, $\propto\langle \hat{E}^{\dag}(\omega)\hat{E}(\omega)\rangle$  for the field component $\mathbf{k}_{\bot}$ (Appendix C).

Inserting the output field, Eq. (\ref{aout}) with $s\rightarrow-$, into Eq. (\ref{Idef}), and expanding to lowest order in $q\hat{z}_n$ (atoms contained in traps), we find that the nonlinear part of the spectrum originates from the correlator  $\langle\hat{z}_n(-\omega)\hat{z}_m(\omega)\rangle$, which is evaluated using the solution (\ref{zo}). Finally, we obtain (Appendix C)
\begin{eqnarray}
I_{\mathbf{k}_{\bot}}(\omega)&=&|r|^2 \frac{|\beta_{\mathbf{k}_{\bot}}|^2}{|\beta_{\mathbf{k}_{\bot}=0}|^2} 2\pi \delta(\omega)
\nonumber\\
&+&|r|^2 32\eta^4\frac{T_e}{E_R}\sum_{jj'}M_{jj'}\frac{\alpha_0}{\nu^2}\chi^{\ast}_j(\omega)\chi_{j'}(\omega).
\label{I}
\end{eqnarray}
The first term is the linear reflection with reflection coefficient
\begin{equation}
r=-\frac{i(\gamma+\Gamma)/2}{i(\gamma+\Gamma)/2+\delta_L-\Delta}.
\label{r}
\end{equation}
At cooperative resonance, $\delta_L=\Delta$, the reflection is perfect \cite{coop}. However, realistically, for the atoms to thermalize inside the traps, we require a small red detuning $0<\Delta-\delta_L\ll \gamma+\Gamma$, which may slightly reduce the reflection [Eq. (\ref{Te}), for finite $T_e>0$].

The second term describes a nonlinear scattered component ($\omega\neq 0$), originated in motion fluctuations inside the traps, who are in turn caused by the light-induced Langevin force $\hat{f}_n(t)$, with an effective temperature $T_e$. Indeed, the frequency dependence of this component derives from the overlap of the collective mechanical responses $\chi_j(\omega)$; its intensity is proportional to $T_e$ and to $\eta^4\propto1/V$, with $\alpha_0=\alpha_{n=0}$ being the friction at the center of the array (atom $n=0$). Since $\chi_j(\omega)$ are centered around the collective mechanical resonances $\nu_j$, this gives rise to sidebands, whose weights are determined by the spatial structure of the modes, contained in the overlap factors
$M_{jj'}$ (see Appendix C, Eq. \ref{Mjj}).

\begin{figure}
\begin{center}
\includegraphics[width=\columnwidth]{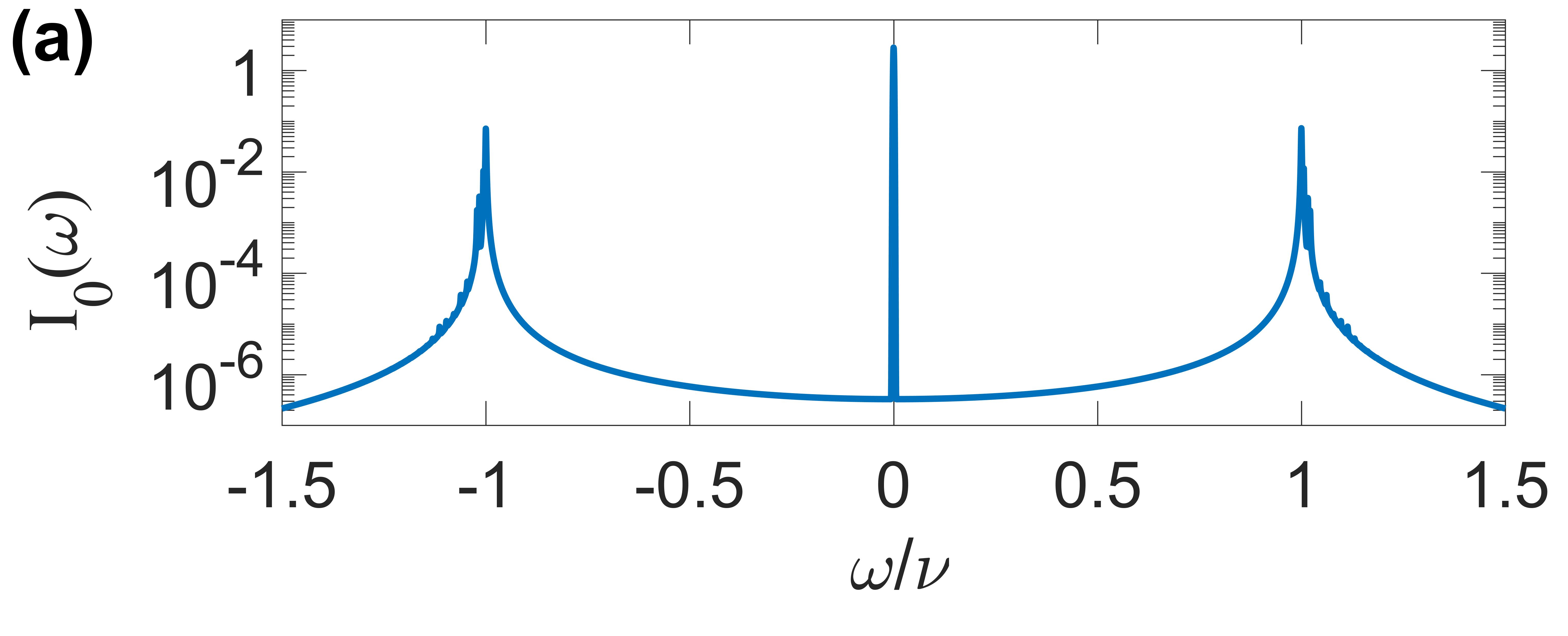}
\includegraphics[scale=0.229]{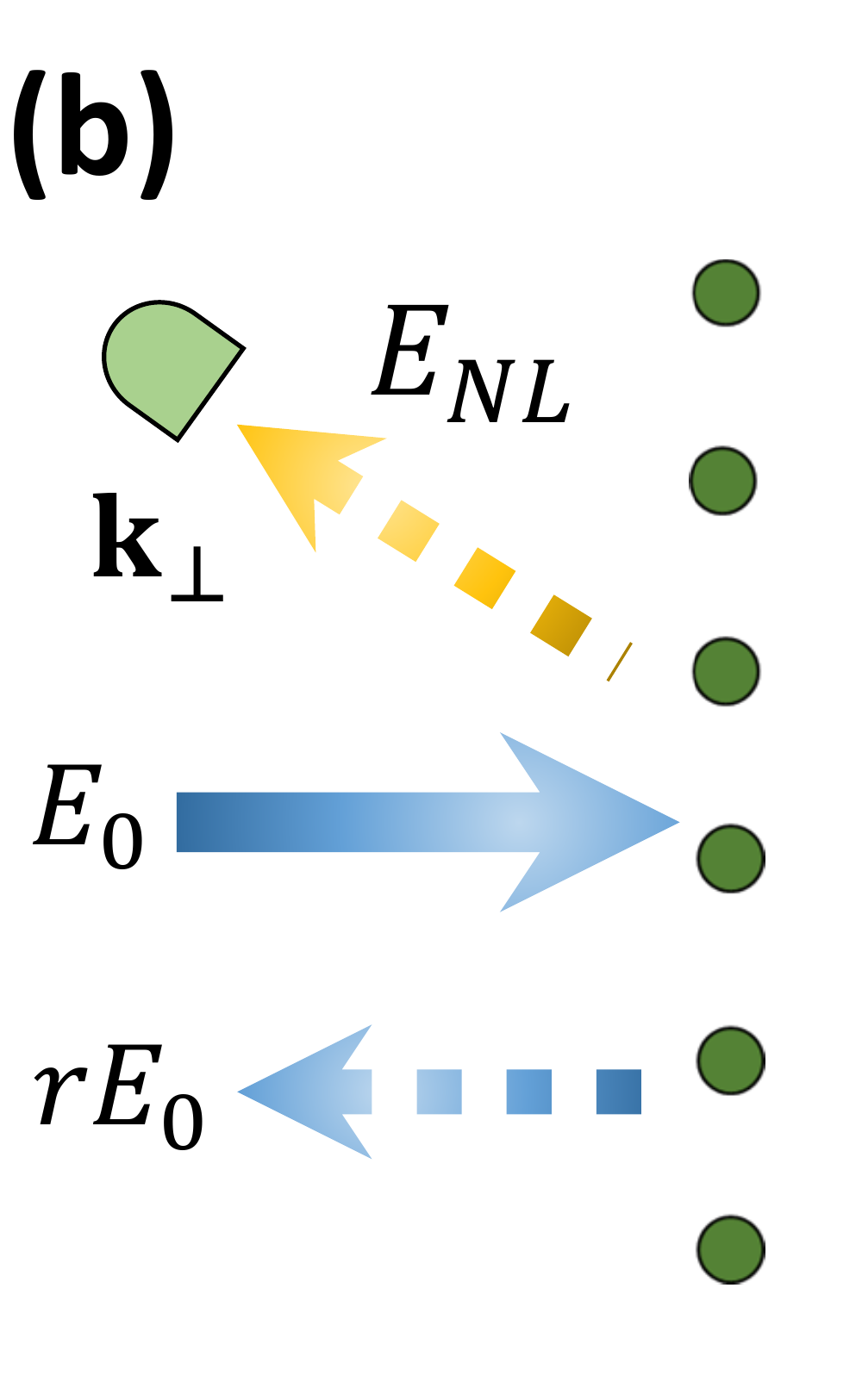}
\includegraphics[scale=0.04]{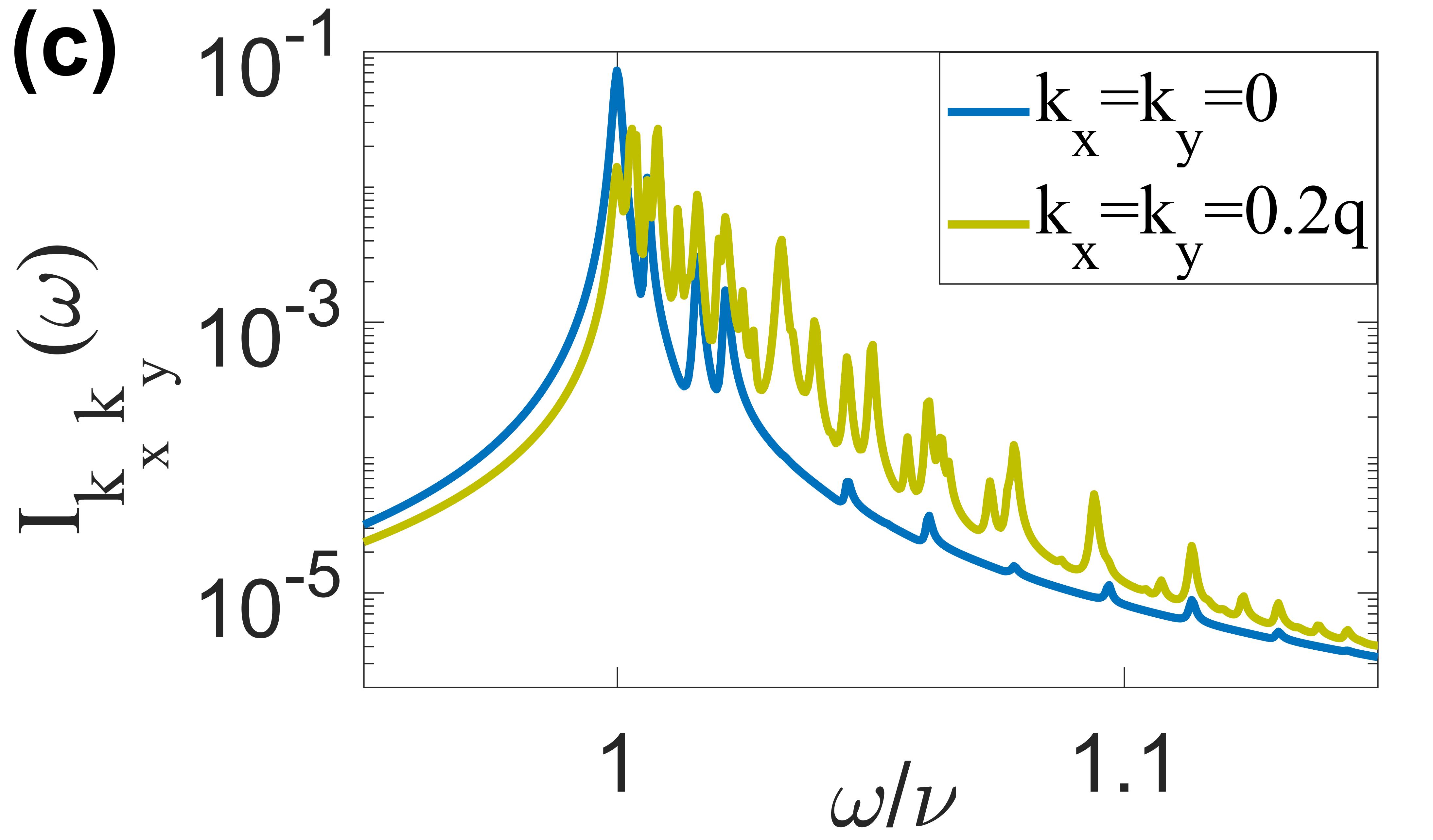}
\caption{\small{
Collective mechanical sidebands in reflected light. (a) Intensity Spectrum of the normal spatial component of the reflected light [Eq. (\ref{I}) for $\mathbf{k}_{\bot}=0$, divided by the experiment time $L/c$]. The central sharp peak at the laser frequency ($\omega=0$) is due to the strong linear reflection $|r|^2$ [with the $\delta(\omega)$  peak approximated here by a Gaussian of width $c/L\rightarrow \alpha_0/3$]. The incident light excites the collective mechanical modes from Fig. 2 which are revealed by the appearance of sidebands centered around their eigenfrequencies $\nu_j$ (same parameters as in Fig. 2, yielding $|r|^2=0.8$). (b) The $|\mathbf{k}_{\bot}|>0$ nonlinear components of the spectrum can be detected at the corresponding far-field angles.
(c) Zoom-in of the sideband for different detection angles: $\mathbf{k}_{\bot}=0$ (blue, same as Fig. 4a), $\mathbf{k}_{\bot}=(0.2q,0.2q)$ (green). Sidebands at higher frequencies $\nu_j$ are associated with mechanical mode profiles of higher spatial frequencies, and therefore become more pronounced for larger detection angles $|\mathbf{k}_{\bot}|$.
 }} \label{fig4}
\end{center}
\end{figure}

Figure 4a, plotted for detection at $\mathbf{k}_{\bot}=0$ with the atom array realization of Fig. 2, clearly exhibits these spectral features. It contains a narrow peak at $\omega=0$ due to the linear reflection, and two broad sidebands centered around the trap frequency $\pm\nu$. Each sideband however, exhibits multiple peaks, with the frequencies $\nu_j$ from Fig. 2b, as clearly seen in the zoom-in, Fig. 4c (blue curve). We recall that higher $\nu_j$ entail higher spatial frequencies in the structure of the collective mode function (Fig. 2c,d). Indeed, Fig. 4c (green curve) displays that by increasing the detection angle, $\mathbf{k}_{\bot}>0$, the sideband components $\nu_j$ beyond $\pm\nu$ become much more prominent.

\section{Quantum squeezing of output light}
Optical nonlinear phenomena, as the one revealed in the previous section, may in general lead to quantum squeezing in the scattered light; namely, to the reduction of its quadrature quantum noise below that of the vacuum, and which is associated with the generation of entangled photon pairs \cite{MW,DF}. The generation of squeezing in the atom-array system can be understood by considering the light-induced motion, $z_n\propto |E|^2$, driven by a field $E$ containing a coherent part $\bar{E}$, and a small vacuum fluctuations component $\hat{\mathcal{E}}$. Then, $z_n\sim (\bar{E}^{\ast}+\hat{\mathcal{E}}^{\dag})(\bar{E}+\hat{\mathcal{E}})\approx |\bar{E}|^2 +\bar{E}^{\ast}\hat{\mathcal{E}}+ \bar{E}\hat{\mathcal{E}}^{\dag}$, so that the phase factor of the output field, $e^{i 2q z_n}$, has the Bogoliubov-transformation form of a squeezed field.
Since the system is inherently multimode along the transverse, array plane, entanglement is generated not only between different longitudinal ($\omega=ck-\omega_L$), but also between different transverse ($\mathbf{k}_{\bot}$) photon modes, giving rise to spatio-temporal quantum squeezing \cite{KOL,LUG}. In this section, we analyze the quantum noise of the output field, and reveal the possibility for nearly perfect quantum squeezing.

To this end, we consider the quantum fluctuations of the reflected field ($s\rightarrow-$) from Eq. (\ref{aout}), assuming uniform illumination, $\Omega_n=\Omega$, for which the collective mechanical modes are lattice Fourier modes, $j\rightarrow\mathbf{k}_{\bot}$, with eigenfrequencies $\nu_{\mathbf{k}_{\bot}}$ and corresponding friction $\alpha_{\mathbf{k}_{\bot}}=\alpha$ (Appendix A).
Working near cooperative resonance, where the reflection $r\approx -1$, and expanding to lowest order in $q\hat{z}_n$ and in the vacuum field (Bogoliubov approximation), we obtain (Appendix D)
\begin{equation}
\widetilde{a}_{\mathbf{k}_{\bot}}(\omega)=u_{\mathbf{k}_{\bot}}(\omega)\hat{a}_{\mathbf{k}_{\bot}}(\omega)+v_{\mathbf{k}_{\bot}\omega}\hat{a}^{\dag}_{-\mathbf{k}_{\bot}}(-\omega),
\label{aout2}
\end{equation}
with the coefficients
\begin{eqnarray}
u_{\mathbf{k}_{\bot}}(\omega)&=&-1+v_{\mathbf{k}_{\bot}}(\omega),
\nonumber\\
v_{\mathbf{k}_{\bot}}(\omega)&=&i8\eta^4\frac{4|\Omega|^2}{(\gamma+\Gamma)^2}\frac{\hbar(\gamma+\Gamma)}{E_R}\frac{\nu^2}{\nu_{\mathbf{k}_{\bot}}^2}\chi_{\mathbf{k}_{\bot}}(\omega).
\label{bog}
\end{eqnarray}
Here the output field fluctuation $\widetilde{a}_{\mathbf{k}_{\bot}}(\omega)\equiv\widetilde{a}_{\mathbf{k}_{\bot}k-}-r\beta\delta_{\mathbf{k}_{\bot}0}\delta_{kq}$ is given in terms of
the incident vacuum fields of the right-propagating modes, $\hat{a}_{\mathbf{k}_{\bot}}(\omega)=\hat{a}_{\mathbf{k}_{\bot},k,+}$ and $\hat{a}_{\mathbf{k}_{\bot}}(-\omega)=\hat{a}_{\mathbf{k}_{\bot},2q-k,+}$, with $\omega=ck-\omega_L$. In general, the output field depends on the vacua of both right- and left-propagating photon modes; however, here we assumed nearly perfect reflection, $r\approx-1$, so that the vacuum fields incident from the right (left-propagating $s\rightarrow-$), are reflected back and do not arrive to the detector at the left. The general case, beyond nearly-perfect reflection, is discussed in Appendix D, and yields similar results.

The output field fluctuations from Eq. (\ref{aout2}) have the typical form of a squeezed vacuum field, whose reduced quantum fluctuations can be measured via homodyne detection, wherein the output field at the correlated angles $\pm\mathbf{k}_{\bot}$ interferes with a strong coherent local oscillator field with phase $\theta$ (Fig. 5a). The relevant fluctuating part of the detected signal is given by the quadrature operator,
$\hat{X}^{\theta}_{\mathbf{k}_{\bot}}(\omega)=e^{-i\theta}\widetilde{a}_{\mathbf{k}_{\bot}}(\omega)+ e^{i\theta}\widetilde{a}^{\dag}_{-\mathbf{k}_{\bot}}(-\omega)$, with the corresponding spatio-temporal noise spectrum,
$S^{\theta}_{\mathbf{k}_{\bot}}(\omega)\equiv \langle \hat{X}^{\theta}_{-\mathbf{k}_{\bot}}(-\omega) \hat{X}^{\theta}_{\mathbf{k}_{\bot}}(\omega)\rangle$ \cite{DF,KOL}.

For each spatio-temporal frequency, $(\mathbf{k}_{\bot},\omega)$, there exists a local oscillator phase that minimizes the noise $S^{\theta}_{\mathbf{k}_{\bot}}(\omega)$ \cite{DF,YP}. The resulting spectrum of minimal noise level, the so-called squeezing spectrum, is given by
 \begin{eqnarray}
S_{\mathbf{k}_{\bot}}(\omega)&=&\left(|u_{-\mathbf{k}_{\bot}}(-\omega)|-|v_{\mathbf{k}_{\bot}}(\omega)|\right)^2
\nonumber\\
&=&\left(\sqrt{|v_{\mathbf{k}_{\bot}}(\omega)|^2+1+2\mathrm{Re}[v_{\mathbf{k}_{\bot}}(\omega)]}-|v_{\mathbf{k}_{\bot}}(\omega)|\right)^2.
\nonumber\\
\label{Sk}
\end{eqnarray}
In the absence of motion, $\chi_{\mathbf{k}_{\bot}}=0$, we have $v_{\mathbf{k}_{\bot}}=0$ and $u_{\mathbf{k}_{\bot}}=-1$, such that the output is just the reflected vacuum with the standard vacuum noise-level $S_{\mathbf{k}_{\bot}}(\omega)=1$. When motion, and hence nonlinearity exist, we may obtain noise reduction (squeezing), $S_{\mathbf{k}_{\bot}}(\omega)<1$.

\begin{figure}
\begin{center}
\includegraphics[scale=0.22]{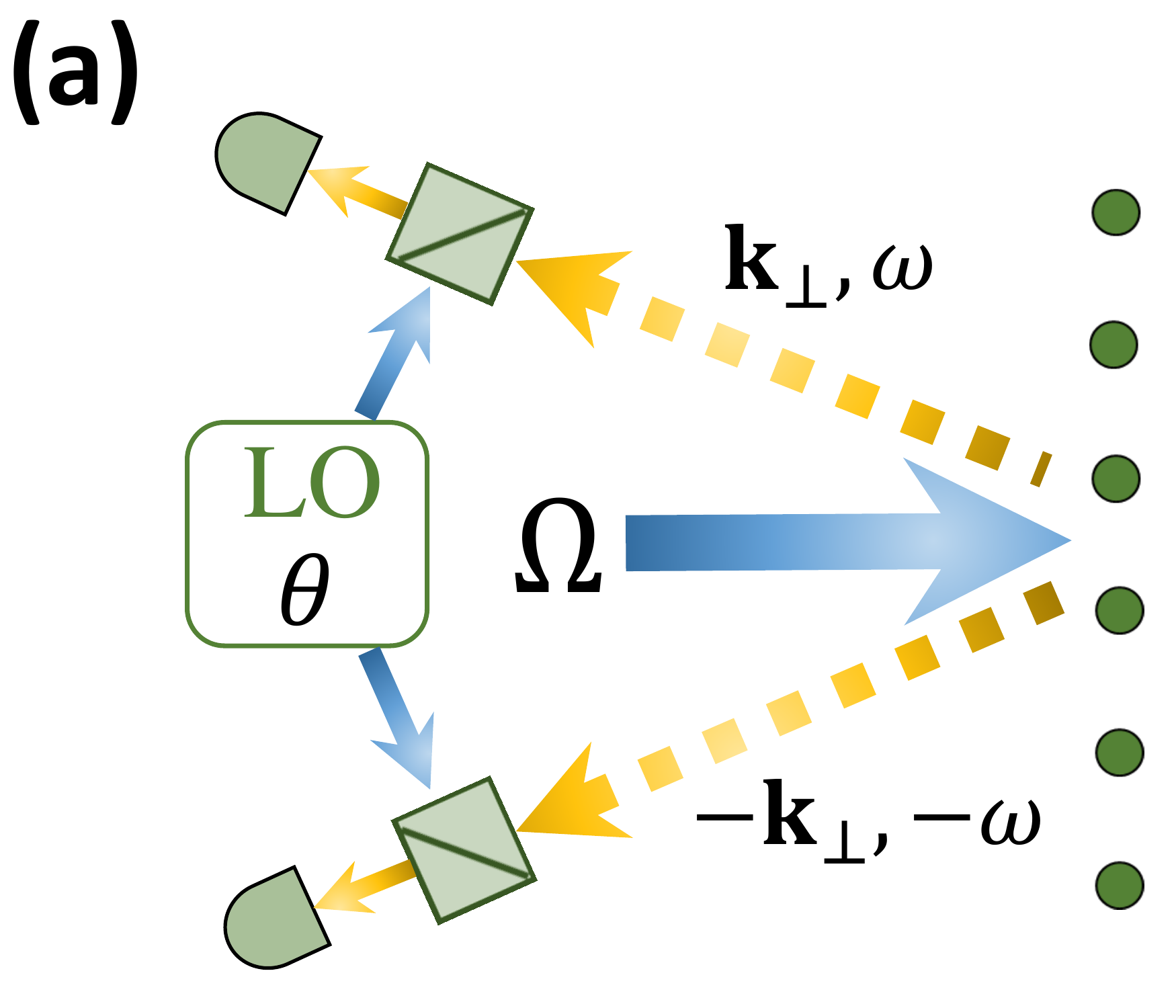}
\includegraphics[scale=0.036]{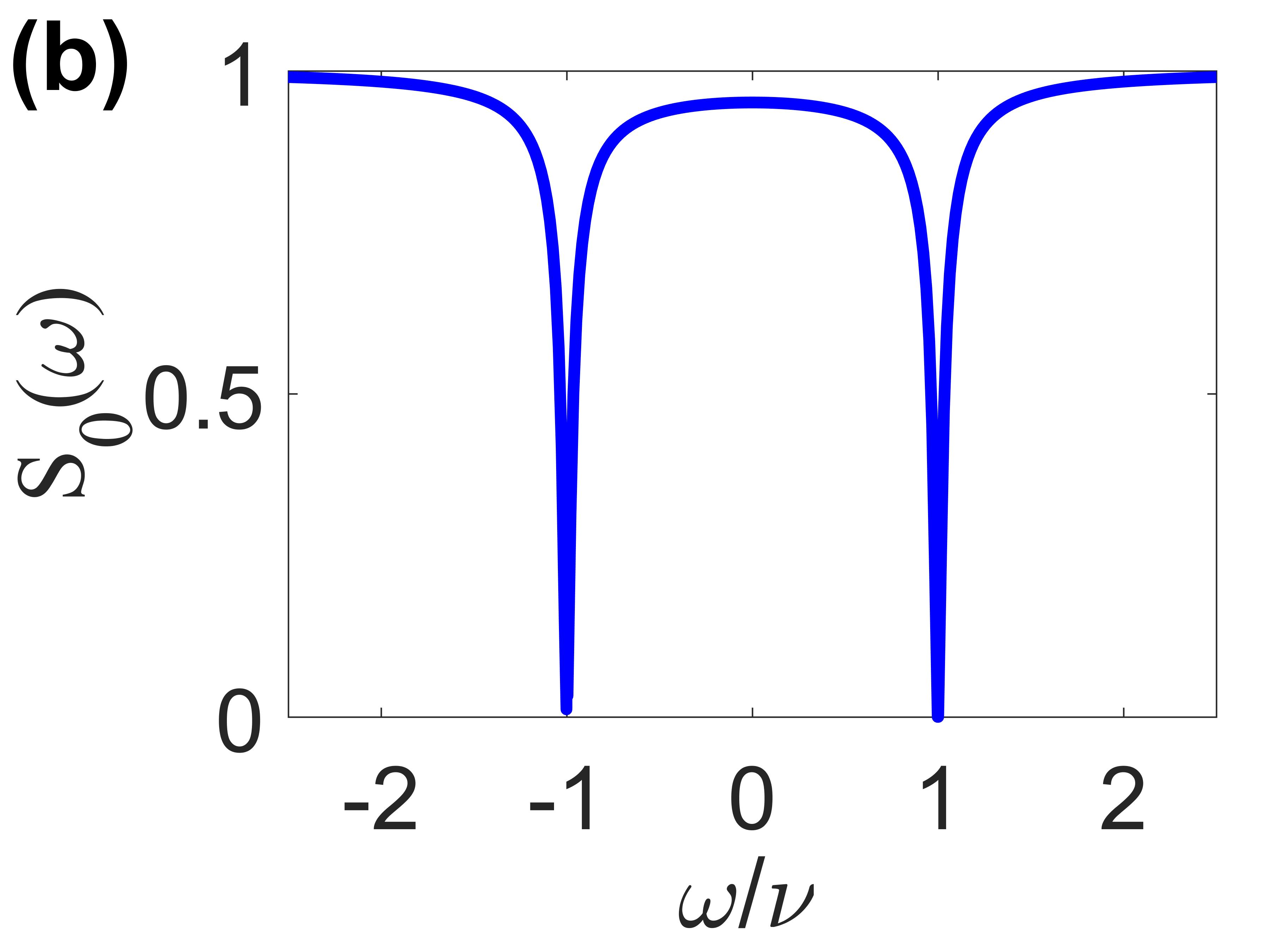}
\includegraphics[scale=0.0355]{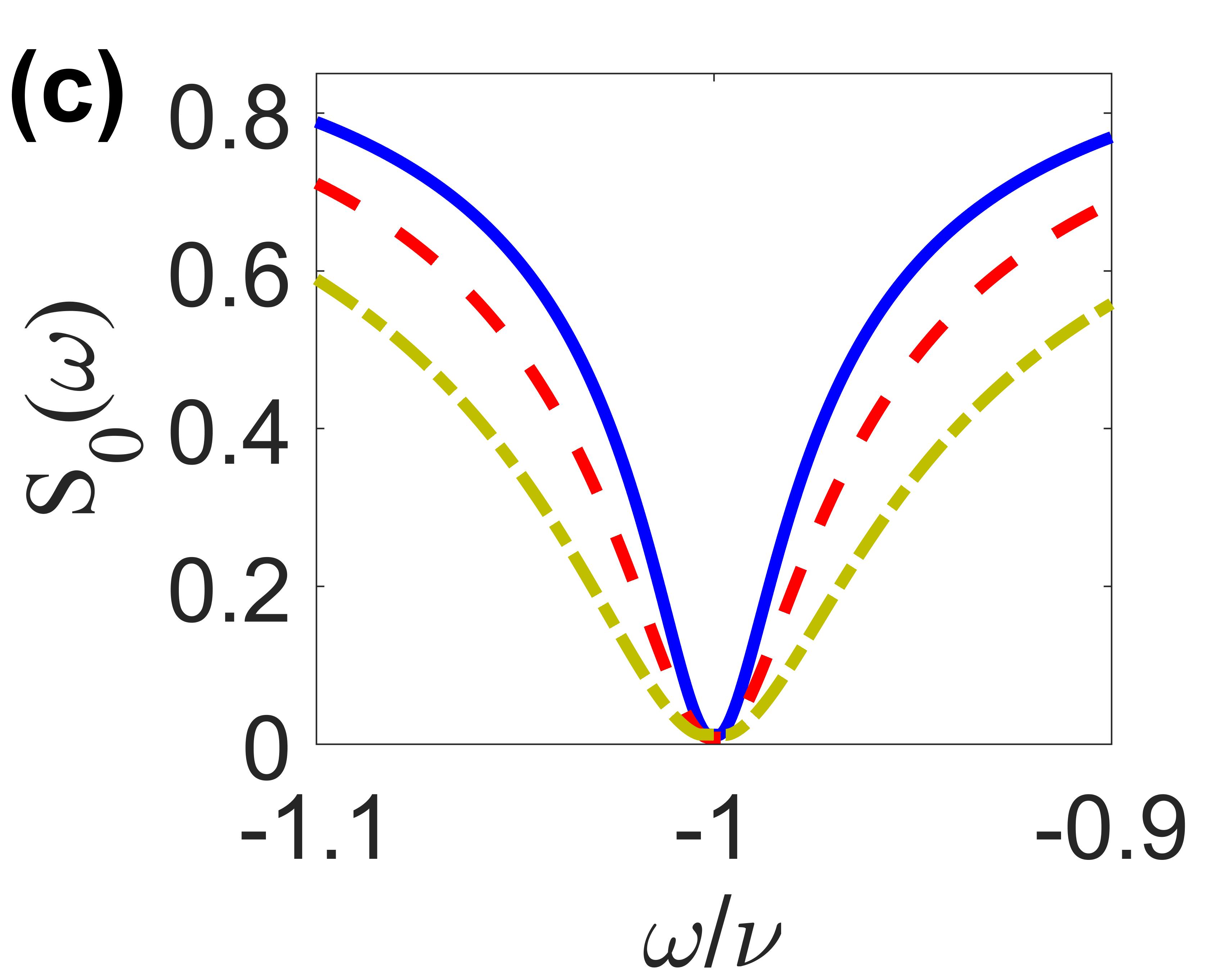}
\includegraphics[scale=0.0355]{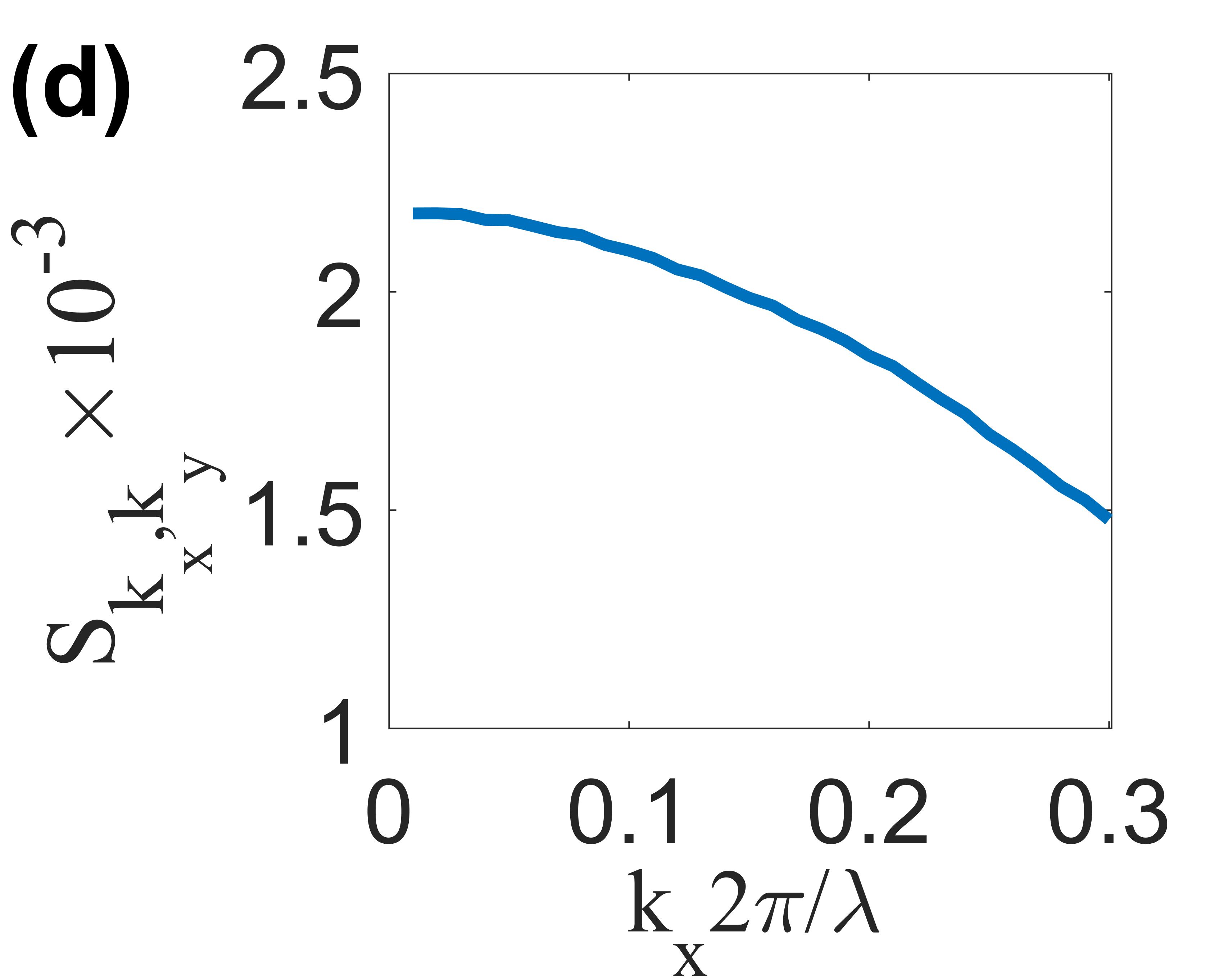}
\caption{\small{
Quantum squeezing of the reflected field. (a) The correlation between spatio-temporal modes $(\pm\mathbf{k}_{\bot},\pm\omega)$ can be observed by homodyne detection at both angles $\pm\mathbf{k}_{\bot}$. (b) Squeezing spectrum, Eq. (\ref{Sk}), as a function of $\omega$ for fixed  $\pm\mathbf{k}_{\bot}=0$. The two dips at the corresponding mechanical resonance $\pm \nu_{\mathbf{k}_{\bot}=0}=\pm\nu$ exhibit very strong squeezing, as expected from Eq. (\ref{peak}) [e.g. reaching $S\sim 10^{-3}$ (see panel d) for the specific parameters considered here: $a/\lambda=0.5$, $V/E_R=1500$, trap length $400$nm (yielding $\eta\approx 0.1$), $\delta_L-\Delta=-0.1(\gamma+\Gamma)$, and
uniform illumination with $\Omega/\gamma=0.1$]. (c) The bandwidth of the squeezing-spectrum dip is determined by the parameter $W_{\mathbf{k}_{\bot}}$ and can be increased by increasing either $a$ or $\Omega$ [Eqs. (\ref{peak}) and (\ref{W})]. This is demonstrated by the red (dashed) curve ($a/\lambda$ increased to $0.6$) and the green (dash-dot) curve  ($\Omega/\gamma$ increased to $0.15$), as compared to the blue curve [same as (b)]. (d) The dependence of the squeezing on $\mathbf{k}_{\bot}$ (spatial squeezing) signifies the quantum correlations generated between the spatial modes $\pm\mathbf{k}_{\bot}$, reflecting the unique multimode and nonlocal optomechanical response of the array. Here this dependence is plotted as a function of $k_x$ for fixed $\omega=\nu-14\alpha$ and $k_y=0$.
 }} \label{fig5}
\end{center}
\end{figure}

\subsection{Squeezing bandwidth and strength}
We observe that maximal squeezing (minimal $S$) is obtained for a large coefficient $|v_{\mathbf{k}_{\bot}}|\gg 1$, i.e. near the collective mechanical resonance $\omega=\pm\nu_{\mathbf{k}_{\bot}}$ where $\chi_{\mathbf{k}_{\bot}}(\omega)$ is very large \cite{comment2}, and where the spectrum (\ref{Sk}) can be approximated as $S_{\mathbf{k}_{\bot}}(\omega)\approx 1/(4|v_{\mathbf{k}_{\bot}}(\omega)|^2)$, yielding
\begin{equation}
S_{\mathbf{k}_{\bot}}(\omega)\approx \frac{1}{W_{\mathbf{k}_{\bot}}^2}\left[\left(\frac{\omega}{\nu_{\mathbf{k}_{\bot}}}\pm1\right)^2+\left(\frac{\alpha}{2\nu_{\mathbf{k}_{\bot}}}\right)^2\right].
\label{peak}
\end{equation}
The quadratic, power-law frequency-dependence means that the bandwidth of the squeezing spectrum near mechanical resonance $\pm \nu_{\mathbf{k}_{\bot}}$ scales as $W_{\mathbf{k}_{\bot}}\nu_{\mathbf{k}_{\bot}}$, with
\begin{equation}
W_{\mathbf{k}_{\bot}}=\frac{\nu^2}{\nu_{\mathbf{k}_{\bot}}^2}B, \quad B=\frac{16|\eta\Omega|^2}{(\gamma+\Gamma)\nu}\Rightarrow \frac{16|\bar{g}|^2}{\kappa\nu}.
\label{W}
\end{equation}
As the parameter $B$ increases, so do the bandwidth and strength of the squeezing, suggesting that this parameter is related to the motion-induced optical nonlinearity. Indeed, by using the mapping to cavity optomechanics from Table I (rightwards double arrow), we note that $B$ is equal to the nonlinear frequency shift of the cavity/resonator, $\sim|\bar{g}|^2/\nu$, in units of its linewidth $\kappa$ \cite{RAB,GIR}. For the atom array, we note that the squeezing bandwidth $\sim B$ can be enlarged by increasing e.g. the Rabi frequency $\Omega$ (avoiding saturation) or the lattice spacing $a$ (recalling that $\gamma+\Gamma\propto \lambda^2/a^2$).

The squeezing strength can become arbitrary large in principle, with a  minimal noise level of order $W_{\mathbf{k}_{\bot}}^{-2}(\alpha/\nu_{\mathbf{k}_{\bot}})^2$. This is typically a very small number, reflecting the mechanical quality factor $\nu/\alpha$ of trapped atoms, or, equivalently, the optomechanical cooperativity $|\bar{g}|^2/(\kappa\alpha)$.

\subsection{Temporal and spatial squeezing spectra}
Fixing the detection angle to $\mathbf{k}_{\bot}=0$, we study the dependence of the squeezing on the frequency $\omega$ (temporal squeezing spectrum) in Fig. 5b. Considering realistic parameters, we observe very strong noise reduction at the corresponding pair of mechanical resonance dips, $\pm \nu_{\mathbf{k}_{\bot}=0}=\pm\nu$, as anticipated above. In Fig. 5c we focus on one of the resonance dips and observe that its bandwidth is much wider than the mechanical linewidth $\alpha/\nu\sim 10^{-4}$ (blue solid curve). Furthermore, it is seen that the squeezing bandwidth widens by increasing either the lattice spacing (red dashed curve) or the Rabi frequency (green dash-dot curve), in agreement with the analysis of Eqs. (\ref{peak}) and (\ref{W}).

Figure 5d displays the dependence of the squeezing on the spatial frequency $k_x$  (spatial squeezing spectrum), by setting $\omega=\nu-14\alpha$ and $k_y=0$. The dependence on $k_x$ is a signature of an effectively \emph{nonlocal} optical nonlinearity of the atom array \cite{KRO,EITNLO}, whose nonlocality is originated in the dipole-dipole interactions between the atoms.

Finally, we address how these results compare with previous studies.
Squeezed light generation via optomechanical nonlinearities was studied theoretically within the standard cavity model \cite{TOM,FAB,CLE} and experimentally with an atom cloud or a membrane inside a cavity \cite{SK3,REG,SAF}. Both the cavity model, and the above analysis of the atom array, predict arbitrary strong squeezing, in principle, with similar scalings of strength and bandwidth (within the bad-cavity regime relevant here) \cite{HAM,CLE}. Interestingly, in our case this is achieved without a cavity and for dozens of atoms (e.g. $\sqrt{N}\sim10\gg 1$). This is in contrast to cavity-confined and macroscopic objects, such as a membrane or thousands of atoms.
More qualitatively, the atom array optomechanics naturally exhibits spatial squeezing, in addition to the temporal squeezing discussed in previous works.

\section{Discussion}
This study establishes the first step in a new direction in the field of quantum optomechanics, wherein the unique collective properties of ordered 2D atomic arrays are harnessed. We stress that the mechanical interactions between the atoms, and the subsequent formation of collective mechanical modes, are not inherent to the atoms, but are rather induced by light (Fig. 2a).
 Our findings demonstrate that 2D atomic arrays exhibit rich and qualitatively new multimode optomechanical phenomena, already at the level of the ``bare", cavity-free system. More quantitatively, the results for squeezing suggest that the optomechanical response of a single mesoscopic atomic array can become comparable or exceed those of current macroscopic cavity systems.

As an extension of this work, one can interface the atomic array with a mirror or cavity. This may offer several advantages. First, the \emph{ordered} array scatters only into the paraxial cavity mode, unlike the case of a disordered cloud \cite{SK3,SK1,SK,CAM}, for which scattering to all directions effectively increases $\kappa$. Second, an atom array inside a cavity or in the vicinity of a mirror, may exhibit a drastic reduction of its radiative linewidth $\gamma+\Gamma$. Considering the mapping to the cavity model, this means that $\kappa$ can become sufficiently small to reach the resolved sideband regime. Combined with the strong optomechanical coupling of the atoms, this may pave the way to observe optomechanical effects at the few-photon level, such as photon-blockade and non-Gaussian states  \cite{RAB,GIR}.

Finally, we point out that ordered atomic arrays were recently proposed as a quantum optical platform enabling various applications, such as tunable scattering properties \cite{coop}, topological quantum photonics \cite{janos,ADM2,janos2}, lasing \cite{ABA}, and enhanced quantum memories and clocks \cite{CHA,ANA,HEN}, all of which based on their collective internal dipolar response. The current study thus opens the way to explore new possibilities by considering and designing both the collective internal and optomechanical responses of atomic arrays.

\acknowledgments
We acknowledge fruitful discussions with Peter Zoller, Dominik Wild, Darrick Chang, Igor Pikovski and J\'{a}nos Perczel, and financial support from the NSF, the MIT-Harvard Center for Ultracold Atoms, and the Vannevar Bush Faculty Fellowship.

\appendix
\section{LIGHT-INDUCED MOTION}
The discussion and derivation of Eq. (\ref{EOM}) is presented in detail in Ref. \cite{notes}. Here, we review the main results leading to this equation, and provide the expressions for its coefficients.

Beginning with the Hamiltonian for photons and atoms and eliminating the photon operators (Markov approximation), we consider the assumptions mentioned in the main text (non-saturated two-level atoms, small-amplitude motion, and paraxial illumination), and obtain \cite{notes}
\begin{eqnarray}
\dot{\tilde{\sigma}}_{n}&=&\left[i(\delta_L-\Delta)-\frac{\gamma+\Gamma}{2}\right]\tilde{\sigma}_{n}+i\sum_{s=\pm}e^{isq\hat{z}_n}\left[\Omega_{ns}+\delta\hat{\Omega}_{ns}(t)\right],
\nonumber\\
\dot{\hat{p}}_n&=&-m\nu^2\hat{z}_n+\hbar q\sum_{s=\pm}\left[i s e^{isq\hat{z}_n}\left(\Omega_{ns}+\delta\hat{\Omega}_{ns}(t)\right)\tilde{\sigma}_n^{\dag}+\mathrm{h.c.}\right]
\nonumber\\
&&+\frac{3}{4}\hbar q\gamma \sum_{m\neq n}\left[\tilde{\sigma}_n^{\dag}F_{nm}q(\hat{z}_n-\hat{z}_m)\tilde{\sigma}_m+\mathrm{h.c.}\right].
\label{HL}
\end{eqnarray}
Here $\tilde{\sigma}_n=\hat{\sigma}_n e^{i\omega_L t}$ is the envelope of the lowering operator of the internal state of atom $n$, $\Omega_{ns}$ is the Rabi frequency of the $s\rightarrow\pm$ propagating incident laser ($s\rightarrow+$ for single-sided illumination), and
\begin{equation}
\delta\hat{\Omega}_{ns}(t)=\sum_{k>0}\sum_{\mathbf{k}_{\bot}}ig_0 e^{i\mathbf{k}_{\bot}\cdot\mathbf{r}^{\bot}_n}e^{-i(ck-\omega_L)t}\hat{a}_{\mathbf{k}_{\bot}ks},
\label{vac}
\end{equation}
is the corresponding Rabi frequency of the vacuum fluctuations [in the paraxial approximation with $g_0=\sqrt{\omega_L/(2\varepsilon_0\hbar A L)}d$]. The cooperative shift and width are given by $\Delta-i\Gamma/2=-(3/2)\gamma\lambda\sum_{n\neq 0} \mathbf{e}_d^{\dag}\cdot\overline{\overline{G}}(\omega_L,\mathbf{r}_{0}^{\bot},\mathbf{r}_{n}^{\bot})\cdot\mathbf{e}_d$ (finding $\Gamma+\gamma=\gamma\frac{3}{4\pi} \frac{\lambda^2}{a^2}$), where $\overline{\overline{G}}(\omega,\mathbf{r},\mathbf{r}')$ is the dyadic Green's function \cite{NH}, $\mathbf{e}_d$ the orientation of the dipole element of the atomic transition (taken as circular polarization) and ``$n=0$" is the atom at the array center.
In the last line, $F_{nm}=\mathbf{e}_d^{\dag}\cdot \overline{\overline{F}}(q\mathbf{r}^{\bot}_n-q\mathbf{r}^{\bot}_m)\cdot\mathbf{e}_d$, $\overline{\overline{F}}$ being the dimensionless tensor
\begin{eqnarray}
&&F_{ij}(\mathbf{u})
\nonumber\\
&&=\delta_{ij}\frac{e^{iu}}{u^2}\left[\left(i-\frac{1}{u}\right)\left(1+\frac{iu-1}{u^2}\right)+\left(\frac{i}{u^2}-2\frac{iu-1}{u^3}\right)\right]
\nonumber\\
&&+\frac{u_i u_j}{u^2}\frac{e^{iu}}{u^2}
\nonumber\\
&&\times\left[\left(i-\frac{3}{u}\right)\left(-1+\frac{3-i3u}{u^2}\right)
+3\left(-\frac{i}{u^2}-2\frac{1-iu}{u^3}\right)\right],
\nonumber\\
%\right.
%\nonumber\\
%&+&\left.3\left(-\frac{i}{a^2}-2\frac{1-ia}{a^3}\right)\right].
\label{F}
\end{eqnarray}
with $i,j\in\{x,y,z\}$, $F_{ij}=\mathbf{e}_i^{\dag}\cdot \overline{\overline{F}}\cdot\mathbf{e}_j$, $u_i=\mathbf{e}_i\cdot \mathbf{u}$ and $u=|\mathbf{u}|$.

By considering the separation of time scales, $\gamma+\Gamma\gg E_R/\hbar,\nu$, the adiabatic steady-state solution of the internal state is found to be \cite{notes},
\begin{equation}
\tilde{\sigma}_n(t)=-\sum_{s=\pm}e^{isq\hat{z}_n}\left[\frac{\Omega_{ns}+\delta\bar{\Omega}_{ns}(t)}{\delta-\Delta+i\frac{\gamma+\Gamma}{2}}+\frac{\Omega_{ns}(sq/m)\hat{p}_n}{\left(\delta-\Delta+i\frac{\gamma+\Gamma}{2}\right)^2}\right],
\label{sig}
\end{equation}
where the last term is a lowest-order correction due to the Doppler effect. Here, $\delta\bar{\Omega}_{ns}(t)\approx \delta\hat{\Omega}_{ns}(t)-\frac{i}{\delta_L-\Delta+i(\gamma+\Gamma)/2}\partial_t\delta\hat{\Omega}_{ns}(t)$ is the vacuum noise in the adiabatic, coarse-grained dynamical picture. The correction, $\partial_t \delta\hat{\Omega}_{ns}$, is required here to guarantee proper quantum dynamics (see below) \cite{notes}.

Inserting Eq. (\ref{sig}) into the equation for $\hat{p}_n$ in (\ref{HL}), to lowest order in $q\hat{z}_n$, we arrive at Eq. (\ref{EOM}) from the main text, with the coefficients (illumination from the left, $\Omega_{ns}=\Omega_n\delta_{s+}$) \cite{notes}:
\begin{eqnarray}
\bar{f}_n&=&\hbar q|\Omega_n|^2\frac{\gamma+\Gamma}{(\delta_L-\Delta)^2+\left(\frac{\gamma+\Gamma}{2}\right)^2},
\nonumber\\
\alpha_n&=&\frac{E_R}{\hbar}|\Omega_n|^2\frac{-2(\delta_L-\Delta)(\gamma+\Gamma)}{\left[(\delta_L-\Delta)^2+\left(\frac{\gamma+\Gamma}{2}\right)^2\right]^2},
\nonumber\\
\hat{f}_n(t)&=&\hbar q\sum_{s=\pm}\left[\frac{i\delta\bar{\Omega}_{ns}^{\dag}\Omega_n+i s\delta\hat{\Omega}_{ns}\Omega^{\ast}_n}{\delta_L-\Delta-i\frac{\gamma+\Gamma}{2}} + \mathrm{h.c.}\right],
\nonumber\\
K_{nm}&=&\frac{3}{4}\hbar q^2\gamma\left[F_{nm}\frac{\Omega_n^{\ast}\Omega_m}{(\delta_L-\Delta)^2+\left(\frac{\gamma+\Gamma}{2}\right)^2} + \mathrm{c.c.}\right].
\nonumber\\
\label{92}
\end{eqnarray}
The Langevin forces $\hat{f}_n(t)$ on different atoms $n,m$ are not independent, since they are originated in the vacuum field and its spatial correlations. Their cross-correlation is found to be
\begin{eqnarray}
&&\langle\hat{f}_n(t)\hat{f}_m(t')\rangle=2D^{nm}_p\delta(t-t')+i2\tilde{D}^{nm}_p\delta'(t-t'),
\nonumber\\
&&D^{nm}_p=(\hbar q)^2\Gamma_{nm}\frac{2\Omega_n^{\ast}\Omega_m}{(\delta_L-\Delta)^2+\left(\frac{\gamma+\Gamma}{2}\right)^2},
\label{fn}
\end{eqnarray}
with $\Gamma_{nm}=3\gamma\lambda \mathrm{Im}[\mathbf{e}_d^{\dag}\cdot \overline{\overline{G}}(\omega_L,\mathbf{r}^{\bot}_n,\mathbf{r}^{\bot}_m)\cdot\mathbf{e}_d]$ being the cooperative decay kernel \cite{LEH}. The term with $\tilde{D}_p^{nm}=-D_p^{nm}\frac{\delta_L-\Delta}{(\delta_L-\Delta)^2+(\gamma+\Gamma)^2/4}$ and $\delta'(t)=\partial_t\delta(t)$, is due to the correction $\partial_t\delta\hat{\Omega}_{ns}(t)$ discussed below Eq. (\ref{sig}), and it guarantees that the dynamics of Eq. (\ref{EOM}) preserve commutation relations and describe genuine quantum Brownian motion \cite{notes,QN}. For the calculations in this paper, however, this correction term is negligible.

For a single atom $n$, $D_p^{nn}$ is the momentum diffusion coefficient \cite{CCT,notes}, which can be associated with an effective temperature of a heat bath formed by the scattering,
\begin{equation}
T_e=\frac{D_p^{nn}}{m\alpha_n}.
\label{Ted}
\end{equation}
Using $D_p^{nn}$ and $\alpha_n$ from Eqs. (\ref{92}) and (\ref{fn}) (noting $\Gamma_{nn}=\gamma$), we obtain $T_e$ from Eq. (\ref{Te}).

Performing the transformation to the collective mechanical modes $j$, $\hat{z}_j=\sum_n U_{jn}^{\ast} \hat{z}_n$, on Eq. (\ref{EOM}), and neglecting the typically very small off-diagonal friction $\alpha_{jj'}\approx \alpha_{j}\delta_{jj'}$ (verified numerically for a variety of incident Gaussian beams), we obtain
\begin{equation}
\dot{\hat{p}}_j=-m\nu^2\hat{z}_j+\bar{f}_j-\alpha_j\hat{p}_j+\hat{f}_j(t),
\quad
\dot{\hat{z}}_j=\hat{p}_j/m.
\label{EOMk}
\end{equation}
For long times $t\gg 1/\alpha_j$ (assuming $\alpha_j>0$, i.e. $\delta_L<\Delta$), the solution in Fourier space, $\hat{z}_n(\omega)=\int_{-\infty}^{\infty}dt e^{i\omega t}\hat{z}_n(t)$, yields Eq. (\ref{zo}). The analysis of the general time-dependent solution is discussed in Ref. \cite{notes}.

Finally, consider the case of uniform illumination, $\Omega_n=\Omega$. The collective mechanical modes $j$ then become 2D lattice Fourier modes, $\mathbf{k}_{\bot}$,
with $\mathbf{k}_{\bot}=(k_x,k_y)$ inside the Brillouin zone associated with the 2D lattice, $k_{x,y}\in\{-\pi/a,\pi/a\}$, and the corresponding
eigenmodes and eigenfrequencies \cite{notes}
 \begin{equation}
\hat{z}_{\mathbf{k}_{\bot}}=\frac{1}{N}\sum_n e^{-i\mathbf{k}_{\bot}\cdot \mathbf{r}_{n}^{\bot}}\hat{z}_n, \quad \nu_{\mathbf{k}_{\bot}}=\sqrt{\nu^2+(K_{\mathbf{k}_{\bot}}-K_0)/m}.
\label{zk}
\end{equation}
Here, $K_{\mathbf{k}_{\bot}}=\sum_{n\neq 0}K_{n0}e^{-i\mathbf{k}_{\bot}\cdot \mathbf{r}_{n}^{\bot}}$ and $\nu_{\mathbf{k}_{\bot}}$ can be evaluated by performing the sum $K_{\mathbf{k}_{\bot}}$ numerically.
The same transformation, $U_{n\mathbf{k}_{\bot}}=(1/\sqrt{N})e^{i\mathbf{k}_{\bot}\cdot\mathbf{r}_n^{\bot}}$, also applies for $\bar{f}_{\mathbf{k}_{\bot}}$ and $\hat{f}_{\mathbf{k}_{\bot}}(t)$. The friction coefficient $\alpha$ is equal to $\alpha_n$ form Eq. (\ref{92}) with $\Omega_n=\Omega$ and is therefore independent of $\mathbf{k}_{\bot}$.

\section{ANALOGY TO CAVITY OPTOMECHANICS}
In the following, we derive the equations of motion for the standard cavity optomechanics model in the linearized regime and discuss the analogy of this model to the atom-array case.

\subsection{Optomechanics in the linearized regime}
Beginning from the linearized Hamiltonian (\ref{Hcl}), we find the equations of motion for the cavity mode $\hat{c}$ and mirror momentum $\hat{p}=im\nu x_0(\hat{b}^{\dag}-\hat{b})$,
\begin{eqnarray}
\dot{\hat{c}}&=&\left[i\delta_c-\frac{\kappa}{2}\right]\hat{c}-i\frac{\bar{g}}{x_0}\hat{z}+ i\delta\hat{\Omega}(t),
\nonumber\\
\dot{\hat{p}}&=&-m\nu^2\hat{z}-2m\nu x_0\left(\bar{g}^{\ast}\hat{c}+\bar{g}\hat{c}^{\dag}\right),
\label{cav1}
\end{eqnarray}
with $\dot{\hat{z}}=\hat{p}/m$ and $\hat{z}=x_0(\hat{b}+\hat{b}^{\dag})$. The cavity damping $\kappa$ and corresponding quantum-noise field $\delta\hat{\Omega}(t)$ are due to the out-coupling from the cavity mirror to outside propagating modes, satisfying $[\delta\hat{\Omega}(t),\delta\hat{\Omega}^{\dag}(t')]=\kappa\delta(t-t')$.

Turning to the atom array, and in analogy to the cavity optomechanics case, we wish to linearize the coupled equations of motion, (\ref{HL}), around the classical steady-state solution. To this end, we consider the classical part of the linear-response solution from (\ref{sig}), $\overline{\sigma}_n=-\frac{\Omega_n}{\delta_L-\Delta+i(\gamma+\Gamma)/2}$, together with $q\hat{z}_n\ll 1$, and write Eqs. (\ref{HL}) to linear order in the operators:
\begin{eqnarray}
\dot{\check{\sigma}}_n&=&\left[i(\delta_L-\Delta)-\frac{\gamma+\Gamma}{2}\right]\check{\sigma}_n-q\Omega_n \hat{z}_n+i\sum_{s=\pm}\delta \hat{\Omega}_{ns}(t),
\nonumber\\
\dot{\hat{p}}_n&=&-m\nu^2\hat{z}_n-\hbar q\left(i\Omega_n^{\ast}\check{\sigma}_n-i\Omega_n\check{\sigma}_n^{\dag}\right)
\nonumber\\
&&+\bar{f}_n+\sum_{m\neq n}K_{nm}\left(\hat{z}_n-\hat{z}_m\right)+\hat{f}_n^{(1)}(t),
\label{atom1}
\end{eqnarray}
where $\check{\sigma}_n(t)=\tilde{\sigma}_n(t)-\overline{\sigma}_n$ is the small amplitude of $\tilde{\sigma}_n(t)$ around its steady-state linear solution $\overline{\sigma}_n$.

The formal equivalence of the equations for $\hat{c}$ and $\check{\sigma}_n$ from (\ref{cav1}) and (\ref{atom1})  is apparent, considering $\bar{g}=-i\eta\Omega_n$ and the rest of the mapping from table I [recalling $\eta=q x_0$ and noting that a phase factor $(-i)$ was dropped in the main text, for simplicity]. This equivalence holds also by comparing the equation for $\hat{p}$ from (\ref{cav1}) with the first line of the equation for $\hat{p}_n$, using $E_R/(\hbar\nu)=2\eta^2$ (we note that a correction to $\nu$ was neglected here). The first term in the second line in  the equation for $\hat{p}_n$ is the average force $\bar{f}_n$ which implicitly exists also in the dynamics for $\hat{p}$ in Eq. (\ref{cav1}), since the latter is written for fluctuations around the average motion [originated in the linearized Hamiltonian (\ref{Hcl})]. The collective mechanical term $K_{nm}$ from the equation for $\hat{p}_n$ does not appear in the Eqs. (\ref{cav1}) for the simple cavity model, however, it can be accounted for by considering a modified cavity model (see subsection 3 below). The last term in Eq. (\ref{atom1}) for $\hat{p}_n$ is a Langevin force, which is absent in the cavity model,
\begin{equation}
\hat{f}_n^{(1)}(t)=\hbar q\sum_{s=\pm}\left[\frac{i s\delta\hat{\Omega}_{ns}\Omega^{\ast}_n}{\delta_L-\Delta-i\frac{\gamma+\Gamma}{2}} + \mathrm{h.c.}\right].
\label{fn1}
\end{equation}
This extra Langevin force originates in the direct coupling between motion and the vacuum field, via the phases $e^{i k_z \hat{z}_n}$ of the photon-atom Hamiltonian. This is in contrast to the cavity model wherein only the cavity mode is directly coupled to the vacuum of the outside modes. This means that for the consideration of quantum noise in the output fields, the two models may not be exactly equivalent.
We note that the force $\hat{f}_n^{(1)}(t)$ appears as a component of the Langevin force from Eq. (\ref{92}).

\subsection{Bad-cavity limit}
We shall now consider the bad-cavity limit, where $\kappa$ is the fastest time scale, and obtain a diffusion equation for the cavity model, in analogy to Eq. (\ref{EOM}). Formally solving the equation for $\hat{c}$ from (\ref{cav1}), within a time interval $\Delta t$ ending at $t$, and denoting the mechanical-mode envelope, $\tilde{b}(t)=e^{i\nu t}\hat{b}(t)$ [recalling $\hat{z}=x_0(\hat{b}+\hat{b}^{\dag})$], we have
\begin{eqnarray}
\hat{c}(t)&=&\hat{c}(t-\Delta t)e^{(i\delta_c-\frac{\kappa}{2})\Delta t}+e^{(i\delta_c-\frac{\kappa}{2}) t}\int_{t-\Delta t}^t dt'e^{-(i\delta_c-\frac{\kappa}{2}) t'}
\nonumber\\
&\times&\left[-i\bar{g}\left(\tilde{b}(t')e^{-i\nu t'}+\tilde{b}^{\dag}(t')e^{i\nu t'}\right)+i\delta\hat{\Omega}(t')\right].
\label{31b}
\end{eqnarray}
Next, we assume the separation of time scales between the fast cavity damping $\kappa$ and the slow mechanical envelope dynamics $\tau_m^{-1}\equiv \dot{\tilde{b}}/\tilde{b}$. This allows to move to coarse-grained dynamics with resolution $\Delta t$ satisfying $\kappa^{-1}\ll \Delta t\ll\tau_m$, where the envelope $\tilde{b}(t')\approx \tilde{b}(t)$ can be pulled outside of the integral, obtaining,
\begin{equation}
\hat{c}(t)\approx \frac{\bar{g} \hat{b}(t)}{\delta_c+\nu+i\kappa/2}+\frac{\bar{g} \hat{b}^{\dag}(t)}{\delta_c-\nu+i\kappa/2}-\frac{\delta\hat{\Omega}(t)}{\delta_c+i\kappa/2}.
\label{31i}
\end{equation}
Here the Langevin noise is taken within a bandwidth $2\pi/\Delta t$ of the coarse-grained time resolution. Finally, inserting this result into the equation for $\hat{p}$ from (\ref{cav1}), we obtain
\begin{equation}
\dot{\hat{p}}\approx-m\nu^2\hat{z}-\alpha_{\mathrm{opt}}\hat{p}+\hat{f}_{\mathrm{opt}}(t),
\label{EOMc}
\end{equation}
where a correction to $\nu$ is neglected here \cite{AKM}. The resulting optically-induced friction and Langevin force read
\begin{eqnarray}
\alpha_{\mathrm{opt}}&=&|\bar{g}|^2 \left[\frac{\kappa}{(\delta_c+\nu)^2+(\kappa/2)^2}+\frac{\kappa}{(\delta_c-\nu)^2+(\kappa/2)^2}\right]
\nonumber\\
&\approx& - |\bar{g}|^2 2\nu\frac{2\delta_c\kappa}{\left[\delta_c^2+(\kappa/2)^2\right]^2},
\nonumber\\
\hat{f}_{\mathrm{opt}}(t)&=&\frac{\hbar}{x_0}\left[\frac{\bar{g}\delta\hat{\Omega}^{\dag}(t)}{\delta_c-i\kappa/2}+\frac{\bar{g}^{\ast}\delta\hat{\Omega}(t)}{\delta_c+i\kappa/2}\right],
\label{32a}
\end{eqnarray}
The second approximate equality in $\alpha_{\mathrm{opt}}$ is valid within the unresolved sideband limit, $\kappa\gg \nu$. Coming back to the condition $\tau_m^{-1}\ll \kappa$ for existence of separation of time scales (and coarse-grained dynamics), we can identify from Eq. (\ref{EOMc}) and the expression for $\alpha_{\mathrm{opt}}$ (e.g. for $\kappa \gtrsim \nu,\delta_c$), that $\tau_m^{-1}\sim \alpha_{\mathrm{opt}}\lesssim |\bar{g}|^2/\kappa$, so that the separation of time scales requires the so-called weak coupling regime, $\kappa\gg \bar{g}$.

Considering the mapping from table I it is easy to verify that the friction coefficients $\alpha_n$ [Eq. (\ref{92}), atom-array model] and  $\alpha_{\mathrm{opt}}$ [Eq. (\ref{32a}), cavity model] are identical within the bad-cavity limit $\kappa\gg \bar{g},\nu$, wherein $\kappa$ is the fastest time scale (in analogy to $\gamma+\Gamma$, the fastest time scale assumed for the atom array).
The analogy between the Langevin forces, $\hat{f}_{\mathrm{opt}}(t)$ from Eq. (\ref{32a}), and $\hat{f}_{n}(t)$ from Eq. (\ref{92}), is apparent if we identify the input vacuum fluctuations $\delta\hat{\Omega}$ with the vacuum field on a single-atom, $\delta\hat{\Omega}_n$. We recall that the absence of an average-force term, $\bar{f}$, in Eq. (\ref{EOMc}) is merely due to the fact that it is already written for fluctuations around the average motion.

\subsection{Collective mechanical coupling}
In order to account for the multimode mechanics of the atom array, we replace the single-mode $\hat{b}$ of the cavity-optomechanics model by the modes $\hat{b}_n$, such that the corresponding mechanical and interaction terms in the Hamiltonian (\ref{Hc}) become $\hbar\nu\sum_n \hat{b}_n^{\dag} \hat{b}_n$ and
$\hbar\sum_n g_n\hat{c}^{\dag}\hat{c}(\hat{b}_n+\hat{b}_n^{\dag})$, respectively, with $g_n$ the optomechanical coupling between the mode $n$ and the cavity. The interaction term in Eq. (\ref{cav1}) for $\hat{c}$ then becomes, $-i\sum_n\bar{g}_n (\hat{b}_n^{\dag}+\hat{b}_n)$, with $\bar{g}_n=g_n\bar{c}$. This results in an equation of motion for the momentum $\hat{p}_n$ of the mechanical mode $n$, in the from of Eq. (\ref{EOMc}), but with an additional interaction term $-\sum_{m\neq n} K'_{nm} \hat{z}_m$. The resulting mechanical coupling coefficient is found to be
\begin{eqnarray}
K'_{nm}&=&\frac{\hbar}{x_0^2}\left[\frac{\bar{g}_n^{\ast}\bar{g}_m}{\delta_c-i\kappa/2}+\frac{\bar{g}_n\bar{g}_m^{\ast}}{\delta_c+i\kappa/2}\right]
\nonumber\\
&\rightarrow& 2\hbar q^2(\delta_L-\Delta) \frac{\Omega_n \Omega_m }{(\delta_L-\Delta)^2+\left(\frac{\gamma+\Gamma}{2}\right)^2},
\label{44b}
\end{eqnarray}
where the expression in the second line is obtained via the mapping $\eta\Omega_n=\bar{g}_n$ and by taking real $\bar{g}_n$. The above coefficient $K'_{nm}$, though not identical to $K_{nm}$ from Eq. (\ref{92}), has a similar structure, suggesting that the multimode cavity optomechanical model can indeed capture the multimode motion of the atom array from Eq. (\ref{EOM}).

\section{INTENSITY SPECTRUM}
Here we provide more details on the derivation of the output field. Eq. (\ref{aout}), and the definition and calculation of the intensity spectrum from Eqs. (\ref{Idef}) and (\ref{I}).

\subsection{Output field}
The formal solution for the paraxial photon modes at time $t$, evolved from initial time $t_0$, is found as usual from the original atom-photon Hamiltonian \cite{notes}, as
\begin{eqnarray}
\widetilde{a}_{\mathbf{k}_{\bot}ks}(t)&=&\widetilde{a}_{\mathbf{k}_{\bot}k s}(t_0)+\sum_n g_0^{\ast}e^{-i\mathbf{k}_{\bot}\cdot\mathbf{r}_n^{\bot}}
\nonumber\\
&&\times\int_{t_0}^{t}dt' e^{i(ck-\omega_L)t'}e^{-isq \hat{z}_n(t')}\tilde{\sigma}_n(t'),
\label{3}
\end{eqnarray}
with $\widetilde{a}_{\mathbf{k}_{\bot}ks}(t)=\hat{a}_{\mathbf{k}_{\bot}ks}(t)e^{ick t}$, and where the field is initially in the vacuum state, $\hat{a}_{\mathbf{k}_{\bot}ks}(t_0)|0\rangle=0$. Since we are interested in a steady-state solution for the fields, we take the initial time $t_0$ to be in the far past, $t_0=-\tau\rightarrow-\infty$, whereas the relevant observation time $t$ is taken at the end of the experiment, $t=\tau\rightarrow \infty$.
Inserting the steady-state solution for $\tilde{\sigma}_n$ from (\ref{sig}) into Eq. (\ref{3}) (neglecting the Doppler correction and taking $\delta\bar{\Omega}_{ns}\approx \delta\hat{\Omega}_{ns}$), and adding the laser input $\beta_{\mathbf{k}_{\bot}}$, we arrive at Eq. (\ref{aout}) from the main text. The laser input is added separately since it was taken here as an external input, which is nevertheless equivalent to considering an initial coherent state.

If we neglect the motion, taking $\hat{z}_n\rightarrow 0$ in Eq. (\ref{aout}), we arrive at the result,
\begin{equation}
\widetilde{a}_{\mathbf{k}_{\bot}ks}=\left(\beta_{\mathbf{k}_{\bot}s}\delta_{kq}+\hat{a}_{\mathbf{k}_{\bot}ks}\right)+r\sum_{s'=\pm}\left(\beta_{\mathbf{k}_{\bot}s'}\delta_{kq}+\hat{a}_{\mathbf{k}_{\bot}ks'}\right).
\label{alin}
\end{equation}
Here, we used the expressions for $r$ and $\delta\hat{\Omega}_{ns}$ [Eqs. (\ref{r}) and (\ref{vac})], together with $\beta_{n}=-i\Omega_{n}/g_0$ and $g_0=\sqrt{\omega_L/(2\varepsilon_0\hbar A L)}d=\sqrt{(c/L)(\gamma+\Gamma)/(2N)}$ (recalling $\gamma+\Gamma=\gamma\frac{3}{4\pi}(\lambda^2/a^2)$ \cite{coop}), and considering an incident field from both sides $s$, for generality.
This result reflects the linear response of the mirror to the input from both sides $s\rightarrow\pm$ (average + vacuum fluctuations), with reflection  and transmission coefficients $r$ and $t=1+r$.

\subsection{Intensity spectrum}
The standard definition of the intensity spectrum is given by the intensity in frequency space,
\begin{eqnarray}
G^{(1)}_{\mathbf{k}_{\bot}s}&=&\langle \hat{E}_{\mathbf{k}_{\bot}s}^{\dag}(\omega)\hat{E}_{\mathbf{k}_{\bot}s}(\omega)\rangle
\nonumber\\
&=&\int dt \int dt' e^{-i\omega(t-t')}\langle \hat{E}_{\mathbf{k}_{\bot}s}^{\dag}(t)\hat{E}_{\mathbf{k}_{\bot}s}(t')\rangle,
\label{G}
\end{eqnarray}
where here the detection of a field propagating in the $\mathbf{k}_{\bot}s$ direction is considered. The general expression for the electric field operator in the paraxial approximation is given by
\begin{equation}
\hat{E}_{\mathbf{k}_{\bot}s}(z,t)=\sum_{k>0}E_V e^{isk z}\hat{a}_{\mathbf{k}_{\bot}ks}(t), \quad E_V=\sqrt{\frac{\hbar kc}{2\varepsilon_0 AL}}.
\label{E}
\end{equation}
The field operator that enters into the spectrum (\ref{G}) however, is that detected far from the atom array, \emph{after} the interaction between the laser pulse, of duration $\sim\tau$, and the atom array is over, i.e. for $t>\tau$. Therefore, for the duration $t-\tau$ after the ``passage time" $\tau$, the field propagates freely, and we can write
\begin{equation}
\hat{a}_{\mathbf{k}_{\bot}ks}(t)=e^{-ick(t-\tau)}\hat{a}_{\mathbf{k}_{\bot}ks}(\tau)=e^{-ick t}\widetilde{a}_{\mathbf{k}_{\bot}ks},
\label{7}
\end{equation}
recalling the notation $\widetilde{a}_{\mathbf{k}_{\bot}ks}=\widetilde{a}_{\mathbf{k}_{\bot}ks}(\tau)=e^{ick \tau}\hat{a}_{\mathbf{k}_{\bot}ks}(\tau)$. Substituting (\ref{7}) for $\hat{a}_{\mathbf{k}_{\bot}ks}(t)$ inside the expression for the field in (\ref{E}), and inserting the latter into the spectrum definition (\ref{G}) [with $\sum_k\rightarrow \frac{L}{2\pi c}\int d\omega$ and $2\pi\delta(ck-ck')=\delta_{kk'}L/c]$, we obtain, $G^{(1)}_{\mathbf{k}_{\bot}s}=|E_V|^2\langle \widetilde{a}_{\mathbf{k}_{\bot}ks}^{\dag}\widetilde{a}_{\mathbf{k}_{\bot}ks}\rangle$, which is identical to the definition from Eq. (\ref{Idef}), up to a normalization factor.

\emph{Calculation of the spectrum from Eq. (\ref{I}).---} Inserting the output field (\ref{aout}) into the spectrum (\ref{Idef}) and expanding to second order in $q\hat{z}_n$, we need to evaluate the correlator $\langle\hat{z}_n(-\omega)\hat{z}_m(\omega)\rangle$. Using the solution (\ref{zo}) for $\hat{z}_n(\omega)$, this requires the calculation of $\langle\hat{f}_n(-\omega)\hat{f}_{m}(\omega)\rangle$, which is found from Eq. (\ref{fn}) as
\begin{eqnarray}
&&\langle\hat{f}_n(-\omega)\hat{f}_{m}(\omega)\rangle=
\nonumber\\
&&2\frac{L}{c} D_p^{nm}\left[1+\frac{\omega(\delta_L-\Delta)}{(\delta_L-\Delta)^2+\left(\frac{\gamma+\Gamma}{2}\right)^2}\right]\approx 2\frac{L}{c} D_p^{nm}.
\nonumber\\
\label{corf}
\end{eqnarray}
The second approximate equality is valid for the frequency bandwidth of our slow dynamics, wherein $\omega\ll \gamma+\Gamma$, and amounts to neglecting the $\propto\delta'(t-t')$ correction in the correlation of $\hat{f}_n(t)$ from Eq. (\ref{fn}). By further neglecting small corrections of order $|r|^2q^2\langle\hat{z}_n^2\rangle$ to the amplitude of the linear spectral peak, we finally obtain the result from Eq. (\ref{I}), with
\begin{equation}
M_{jj'}=\frac{\tilde{\beta}_{\mathbf{k}_{\bot}j}\beta_{\mathbf{k}_{\bot}j'}}{|\beta_{\mathbf{k}_{\bot}=0}|^2}\frac{\nu^4}{\nu_j^2\nu{j'}^2}\sum_{n,m}U_{jn}^{\ast}U_{j'm}\frac{\Gamma_{nm}}{\gamma}\frac{\Omega_n^{\ast}\Omega_m}{|\Omega_0|^2},
\label{Mjj}
\end{equation}
and where $\beta_{\mathbf{k}_{\bot}j}=(1/N)\sum_ne^{-i\mathbf{k}_{\bot}\cdot\mathbf{r}_n^{\bot}}U_{jn}\beta_n$, and $\tilde{\beta}_{\mathbf{k}_{\bot}j}=(1/N)\sum_ne^{i\mathbf{k}_{\bot}\cdot\mathbf{r}_n^{\bot}}U_{jn}\beta_n^{\ast}$.

\section{QUANTUM SQUEEZING}
In the following, we elaborate on several topics related to the analysis of the quantum squeezing from Sec. V.

\subsection{Output field fluctuations}
In order to arrive at Eq. (\ref{aout2}) for the quantum fluctuations of the output field, we first expand Eq. (\ref{aout}) to lowest order in $q\hat{z}_n$. Next, we neglect the term proportional to the product of the motion and field fluctuations, $\propto \hat{f}_n \delta\hat{\Omega}_n$, since it is second order in the vacuum fluctuations (Bogoliubov-like approximation/linearization). By considering uniform illumination, $\beta_s=|\beta_s|e^{i\phi_s}$ (from both sides of the array $s\rightarrow\pm$), we then obtain
\begin{eqnarray}
\widetilde{a}_{\mathbf{k}_{\bot}ks}&=&
\left(\beta_s \delta_{\mathbf{k}_{\bot}0}\delta_{kq}+\hat{a}_{\mathbf{k}_{\bot}ks}\right)+r\sum_{s'=\pm}\left(\beta_{s'} \delta_{\mathbf{k}_{\bot}0}\delta_{kq}+\hat{a}_{\mathbf{k}_{\bot}ks'}\right)
\nonumber\\
&-&ir\sum_{s'=\pm}\left[\tilde{\mu}_{\mathbf{k}_{\bot}kss'}\hat{a}^{\dag}_{-\mathbf{k}_{\bot},2q-k,s'}+\bar{\mu}_{\mathbf{k}_{\bot}kss'}\hat{a}_{\mathbf{k}_{\bot},k,s'}\right],
\nonumber\\
\label{42}
\end{eqnarray}
with
\begin{eqnarray}
\tilde{\mu}_{\mathbf{k}_{\bot}kss'}&=&\eta^2\frac{\nu^2}{\nu^2_{\mathbf{k}_{\bot}}}\chi_{\mathbf{k}_{\bot}k}\sum_{pp'}\frac{\beta_p\beta^{\ast}_{p'}c}{L N\nu}(s-p)(p'r^{\ast}+s'r)e^{i2\phi_{p'}},
\nonumber\\
\bar{\mu}_{\mathbf{k}_{\bot}kss'}&=&\eta^2\frac{\nu^2}{\nu^2_{\mathbf{k}_{\bot}}}\chi_{\mathbf{k}_{\bot}k}\sum_{pp'}\frac{\beta_p\beta^{\ast}_{p'}c}{L N\nu}(s-p)(p'r+s'r^{\ast}),
\label{43}
\end{eqnarray}
and where we denoted $\chi_{\mathbf{k}_{\bot}k}=\chi_{\mathbf{k}_{\bot}}(kc-\omega_L)$. The first line is the linear mirror response [Eq. (\ref{alin})], whereas the nonlinear, motion-induced response is described by the second line, which contains the Bogoliubov-type coupling between annihilation and creation field operators. For illumination only from the left ($\beta_s=\beta \delta_{s+}$), the above expression for the reflected field ($s\rightarrow-$) becomes [using $|\beta|^2 c/(LN)=2|\Omega|^2/(\gamma+\Gamma)$],
\begin{equation}
\widetilde{a}_{\mathbf{k}_{\bot}ks}=r\beta\delta_{\mathbf{k}_{\bot}0}\delta_{kq}+\sum_{s=\pm}\left[u_{\mathbf{k}_{\bot}ks}\hat{a}_{\mathbf{k}_{\bot},k,s}+v_{\mathbf{k}_{\bot}ks}\hat{a}^{\dag}_{-\mathbf{k}_{\bot},2q-k,s}\right]
\label{47a}
\end{equation}
with
\begin{eqnarray}
&&u_{\mathbf{k}_{\bot}k+}=r+ir'\mu_{\mathbf{k}_{\bot}k}, \quad\quad\quad  v_{\mathbf{k}_{\bot}k+}=ir'e^{i2\phi}\mu_{\mathbf{k}_{\bot}k},
\nonumber\\
&&u_{\mathbf{k}_{\bot}k-}=1+r-ir''\mu_{\mathbf{k}_{\bot}k}, \quad  v_{\mathbf{k}_{\bot}k-}=r''e^{i2\phi}\mu_{\mathbf{k}_{\bot}k},
\nonumber\\
\label{47b}
\end{eqnarray}
where $r'=\mathrm{Re}[r]$, $r''=\mathrm{Im}[r]$, and $\mu_{\mathbf{k}_{\bot}k}=-ir v_{\mathbf{k}_{\bot}}(\omega)$ with $v_{\mathbf{k}_{\bot}}(\omega)$ from Eq. (\ref{bog}).

At cooperative resonance, $\delta_L=\Delta$, we have $r'=r=-1$ and $r''=0$, so that $u_{\mathbf{k}_{\bot}k-},v_{\mathbf{k}_{\bot}k-}=0$ and the output field depends only on the $s\rightarrow+$ fluctuations. However, in practice, for the atoms to thermalize, we need a non-vanishing friction $\alpha>0$, which requires $\delta_L-\Delta<0$ [Eq. (\ref{92})]. In the main text, we simplify the presentation by considering the regime $|\delta_L-\Delta|\ll\gamma+\Gamma$ for which $r'\approx r \approx -1$ and $r''\ll 1$, taking $r''\rightarrow 0$ in Eq. (\ref{47b}), thus obtaining the field fluctuations in Eq. (\ref{aout2}) and the resulting nearly-perfect squeezing. Allowing for a finite value for $r''$ and $1-r$, leads to extra noise inserted by the vacuum modes $s\rightarrow-$ transmitted from the right, which may slightly degrade the squeezing. This can be avoided however, by considering a modified detection scheme, as discussed in subsection 3 below.

\subsection{Squeezing at mechanical resonance: Discussion}
The analysis of the squeezing around the mechanical resonance $\omega=\pm\nu_{\mathbf{k}_{\bot}}$ in Sec. V [Eqs. (\ref{peak}), (\ref{W}) and Fig. 5c], revealed that its bandwidth is typically much greater than the mechanical width $\alpha$. Therefore, the value of the squeezing exactly on mechanical resonance (e.g. within a width $\alpha$ around it) is unimportant, and the expression from (\ref{peak}) suffices to discuss the squeezing at the resonance for any practical purpose.

Nevertheless, and from a purely formal aspect, we now briefly elaborate on the quantum noise of the field within a width $\alpha$ from $\pm\nu_{\mathbf{k}_{\bot}}$, where Eq. (\ref{peak}) is supposedly invalid (since $\mathrm{Im}[\chi_{\mathbf{k}_{\bot}}(\omega)]$ is large, see comment \cite{comment2}). We first note the commutation relation of the output field from Eq. (\ref{aout2}): $[\widetilde{a}_{\mathbf{k}_{\bot}}(\omega),\widetilde{a}^{\dag}_{\mathbf{k}_{\bot}}(\omega)]=1-2\mathrm{Re}[v_{\mathbf{k}_{\bot}}(\omega)]$. This expression is equal to $1$, as it should, for all $\omega$ except at a  region of width $\sim \alpha$ around the mechanical resonance, where $\mathrm{Re}[v_{\mathbf{k}_{\bot}}(\omega)]\propto\mathrm{Im}[\chi_{\mathbf{k}_{\bot}}(\omega)]$ does not vanish. Formally, this means that any statement on quantum noise at this narrow (and practically irrelevant) region is meaningless, since the commutation relations are wrong. This is an artifact of the adiabatic-elimination (coarse-grained dynamics) we employed, where high frequencies of quantum noise are ignored. In principle, this can be fixed by using a more careful treatment of the output field \cite{optomech}.

\begin{figure}
\begin{center}
\includegraphics[width=\columnwidth]{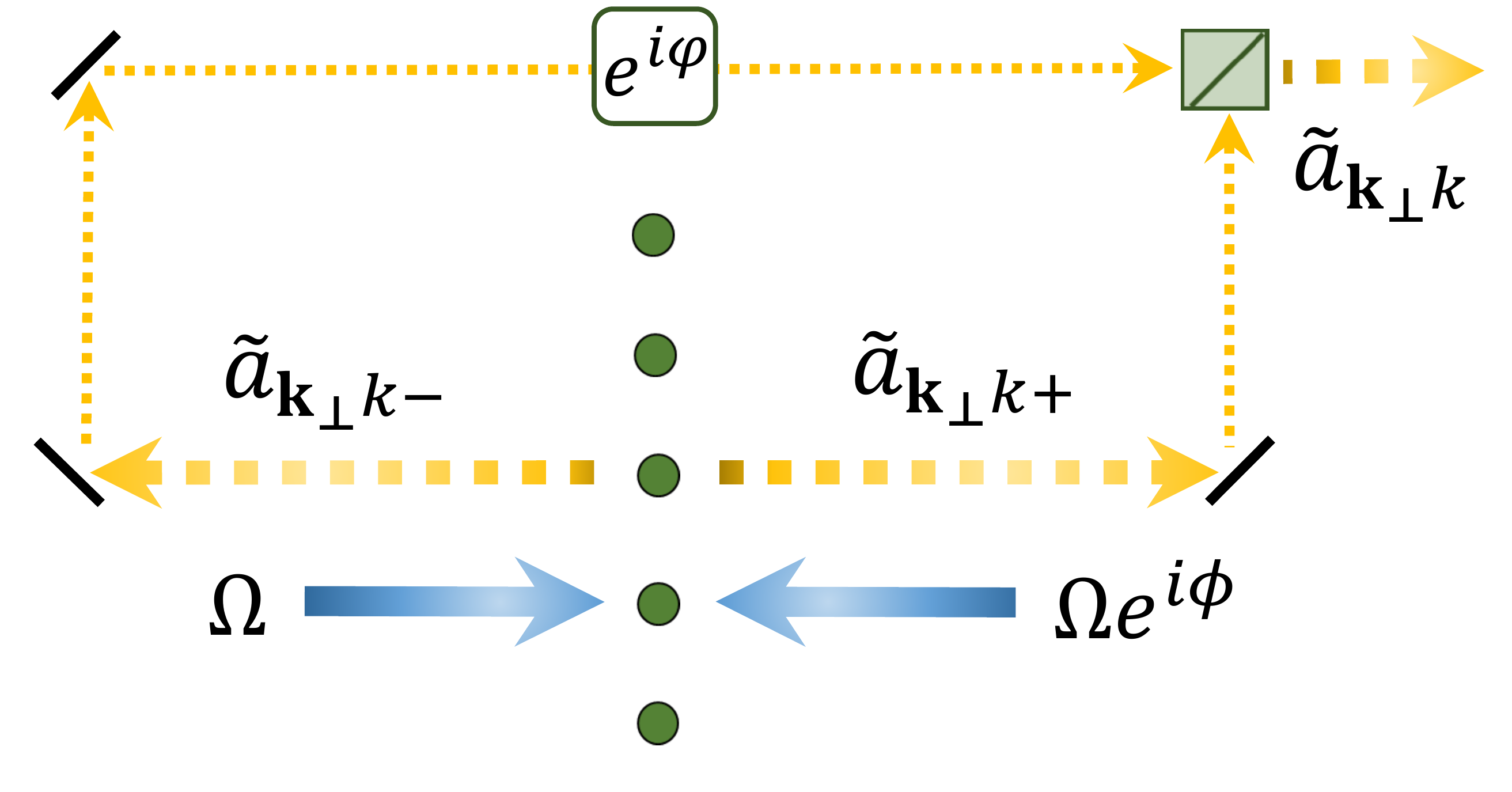}
\caption{\small{
Balanced scheme for the generation and detection of quantum optical squeezing beyond the nearly-perfectly reflecting case. Laser drive is incident from both sides with equal magnitude $\Omega$ and a phase difference $\phi$. The detected output field,  $\widetilde{a}_{\mathbf{k}_{\bot}k}$, is a superposition of the outputs fields from both sides, with a adjustable interference phase $\varphi$.
 }} \label{fig6}
\end{center}
\end{figure}

\subsection{Beyond perfect reflection}
As explained above, the output field (\ref{aout2}) and the resulting squeezing discussed in the main text, are obtained from the more general result of Eq. (\ref{43}), by assuming single-sided illumination and nearly perfect reflection. The latter assumption amounts to $|\delta_L-\Delta|\ll\gamma+\Gamma$, and is used above to neglect the influence of the left-propagating vacuum. This way, one remains with a single output port ($s\rightarrow-$) and a single input port ($s\rightarrow+$), avoiding an additional input port whose noise can degrade the squeezing. Nevertheless, we demonstrate in the following, that even if one gives up nearly-perfect reflection, such that two input ports with their vacuum noises exist, the same optimal squeezing can be achieved by considering a balanced scheme with two output ports \cite{VUL}.

To this end, we consider the scheme from Fig. 6: A uniform incident laser propagates from both sides, with equal magnitude $\Omega$, and a phase difference $\phi$. The detected output field is given by a superposition of the outputs from both sides, $\widetilde{a}_{\mathbf{k}_{\bot}k}=\frac{1}{\sqrt{2}}[\widetilde{a}_{\mathbf{k}_{\bot}k+}+e^{i\varphi}\widetilde{a}_{\mathbf{k}_{\bot}k-}]$, with an adjustable interference phase $\varphi$. Choosing $\phi=\pi$ and $\varphi=0$, and using the expression for the output fields $s\rightarrow\pm$ from Eq. (\ref{43}), we obtain the detected output field fluctuations (subtracting the average),
\begin{equation}
\widetilde{a}_{\mathbf{k}_{\bot}k}=u_{\mathbf{k}_{\bot}k}\check{a}_{\mathbf{k}_{\bot}k}+v_{\mathbf{k}_{\bot}k}\check{a}^{\dag}_{-\mathbf{k}_{\bot},2q-k},
\label{10a}
\end{equation}
with $\check{a}_{\mathbf{k}_{\bot}k}=\frac{1}{\sqrt{2}}[\hat{a}_{\mathbf{k}_{\bot}k+}+\hat{a}_{\mathbf{k}_{\bot}k-}]$ (satisfying $[\check{a}_{\mathbf{k}_{\bot}k},\check{a}^{\dag}_{\mathbf{k}_{\bot}k}]=1$),
%\begin{equation}
%\check{a}_{\mathbf{k}_{\bot}k}=\frac{1}{\sqrt{2}}[\hat{a}_{\mathbf{k}_{\bot}k+}+\hat{a}_{\mathbf{k}_{\bot}k-}, \quad [\check{a}_{\mathbf{k}_{\bot}k},\check{a}^{\dag}_{\mathbf{k}_{\bot}k}]=1,
%\label{10b}
%\end{equation}
and the Bogoliubov coefficients
\begin{eqnarray}
u_{\mathbf{k}_{\bot}k}&=&1+2r+\frac{r}{r^{\ast}}v_{\mathbf{k}_{\bot}k}
\nonumber\\
v_{\mathbf{k}_{\bot}k}&=&i|r|^2 8\eta^4\frac{4|\Omega|^2}{(\gamma+\Gamma)^2}\frac{\hbar(\gamma+\Gamma)}{E_R}\frac{\nu^2}{\nu_{\mathbf{k}_{\bot}}^2}\chi_{\mathbf{k}_{\bot}k}.
\label{10c}
\end{eqnarray}
This output field has the same form as that from Eq. (\ref{aout2}), with the vacuum of the superposition mode $\check{a}_{\mathbf{k}_{\bot}k}$ in the former, replacing that of the $\hat{a}_{\mathbf{k}_{\bot}k+}$ mode in the latter. For $r\rightarrow -1$, the Bogoliubov coefficients in (\ref{10c}) become identical to those from Eq. (\ref{bog}), this time without the need to ignore the noise from any input port. Moreover, even for smaller $|r|$, $|v_{\mathbf{k}_{\bot}k}|$ can still get very large and lead to nearly-perfect squeezing as before.


\begin{thebibliography}{}
\bibitem{comment2} A more precise condition appears to be $|v_{\mathbf{k}_{\bot}}|\gg 1,\mathrm{Re}[v_{\mathbf{k}_{\bot}}(\omega)]$, so that strong squeezing should be observed for frequencies where $\mathrm{Re}[\chi_{\mathbf{k}_{\bot}}(\omega)]\gg \mathrm{Im}[\chi_{\mathbf{k}_{\bot}}(\omega)]$, i.e. close to the mechanical resonance, but slightly detuned (e.g. by a mechanical linewidth $\alpha$). This subtle point has no practical importance, however, considering that the squeezing bandwidth is much wider than $\alpha$ [see Eqs. (\ref{peak}) and (\ref{W})], as further discussed in Appendix D.
\end{thebibliography}
\end{document}